\documentclass[11pt]{amsart}
\usepackage{amsmath,amsfonts,amscd,amssymb,epsf}
\newtheorem{theorem}{Theorem}[section]
\newtheorem{proposition}[theorem]{Proposition}
\newtheorem{lemma}[theorem]{Lemma}

\theoremstyle{definition}

\theoremstyle{remark} \newtheorem{remark}[theorem]{Remark}
\numberwithin{equation}{section}
\DeclareMathOperator{\ad}{ad} \DeclareMathOperator{\Res}{Res}
\DeclareMathOperator{\fixed}{fixed}
\DeclareMathOperator{\dToda}{dToda}

\newcommand{\Z}{{\mathbb{Z}}}

\newcommand{\mL}{\mathcal{L}}
\newcommand{\M}{\mathcal{M}}
\newcommand{\mP}{\mathcal{P}}
\newcommand{\B}{\mathcal{B}}

\newcommand{\pa}{\partial}

\newcommand{\wh}{\widehat}

\newcommand{\vep}{\varepsilon}

\newcommand{\ma}[4]{\begin{pmatrix}
              #1 & #2 \\ #3 & #4
             \end{pmatrix}}
\newcommand{\ve}[2]{\begin{pmatrix}
              #1  \\ #2
             \end{pmatrix}}
\begin{document}
\title{On Dispersionless coupled modified KP hierarchy}
\author{Lee-Peng Teo}
\address{Department of Applied Mathematics \\
National Chiao Tung University, 1001 \\ Ta-Hsueh Road, Hsinchu
City, 30050 \\ Taiwan, R.O.C.} \email{lpteo@math.nctu.edu.tw}
\begin{abstract}
 We define and study dispersionless coupled modified
KP hierarchy, which incorporates two different versions of
dispersionless modified KP hierarchies. \end{abstract}

\maketitle
\section{Introduction}
Recently, the dispersionless limit of integrable hierarchies is
under active research (see, e.g. \cite{K2, K1, TT2, TT5, TT4,
TT1}). There are various problems associated with dispersionless
KP (dKP) and dispersionless Toda (dToda) hierarchies, such as
topological field theory and its connection to string theory,
$2$D-gravity, matrix models and conformal maps (see, e.g.
\cite{TT3, TT6, D1, D2, BX1, BX2, BS, WZ, KKMWZ, BMRWZ, Z, MWZ}).
In contrast, dispersionless modified KP (dmKP) hierarchy is less
under spot light. We found at least two different versions of dmKP
hierarchies, one is considered by Kupershmidt in \cite{Kuper} and
later by Chang and Tu in \cite{CT}, the other is defined by Takebe
in \cite{T1}. It is well known that a solution of dToda hierarchy
will give a solution of dKP hierarchy (\cite{TT1}). In \cite{CT},
Chang and Tu proved that under a Miura map, a solution of dKP
hierarchy will give rise to a solution to their dmKP hierarchy. In
fact, this process can be reversed and we can view the dmKP
hierarchy as a transition between a dToda to a dKP hierarchy. One
of the problem that is still left open is the existence of tau
function for the dmKP hierarchy. In order that a satisfactory tau
function exists, we find that it is necessary to introduce an
extra flow to the dmKP hierarchy considered by Kupershmidt, Chang
and Tu. This is exactly what Takebe did. Takebe's version of dmKP
hierarchy is the dKP hierarchy with an extra flow. Hence we
incorporate the two versions of dmKP hierarchy. We consider a dmKP
hierarchy in Chang and Tu's version, with an extra time parameter,
and call it the dispersionless coupled modified KP hierarchy
(dcmKP). We develop the theory along the lines of Takasaki and
Takebe \cite{TT1} and Takebe \cite{T1}.

The basic object in our dcmKP hierarchy is a formal power series
$\mL$ and a polynomial $\mP$ in variable $k$ with coefficients
functions of time variables $t=(x, s, t_1, t_2, \ldots)$. We
define the hierarchy in terms of Lax equations. We introduce the
dressing function in Section 2. By means of the dressing function,
we define the Orlov function $\M$, which form a canonical pair
with $\mL$, namely $\{\mL, \M\}=1$. We define the $S$ function as
a primitive of a closed one form. The tau function is defined so
that it generates the coefficients of $\M$.

In Section 3, we prove that a solution of the dToda hierarchy give
a solution to our dcmKP hierarchy with $\mP=k$. In Section 4, we
discuss the Miura map which transform a solution of our dcmKP
hierarchy to a solution of Takebe's dmKP hierarchy (dKP hierarchy
with an extra parameter s). We find our Miura map the inverse of
the one considered by Chang and Tu \cite{CT} when $s$ is
considered as a parameter.

In Section 5, we consider the twistor construction of solutions to
our dcmKP hierarchy. We also show that every solution of our dcmKP
hierarchy has an associated twistor data. In Section 6, we
consider the $w_{1+\infty}$ symmetry generated by the action of a
Hamiltonian vector field to the twistor data.

\section{Dispersionless modified KP hierarchy}
\subsection{Lax formalism}
We define dispersionless coupled modified KP hierarchy (dcmKP) by
incorporating the definitions of dispersionless modified KP
hierarchy (dmKP) of \cite{Kuper, CT} and \cite{T1}. The
fundamental quantity
\begin{align*}
\mL= k + \sum_{n=0}^{\infty} u_{n+1}(t) k^{-n}
\end{align*}
is a formal power series in $k$ with coefficients $u_{n+1}(t)$
depend on infinitely many continuous variables $t = (x, s, t_1,
t_2, \ldots,)$. We also introduce an auxiliary monic polynomial of
degree $N$
\[
\mP = k^N + p_{N-1}(t) k^{N-1} + \ldots + p_0(t).
\]
The differential equations that govern the deformation of $\mL$
with respect to $s, t_1, t_2, \ldots, $ are
\begin{align}\label{equation}
\frac{\pa\mL}{\pa t_n} =& \{\B_n, \mL\}, \hspace{1cm} \B_n
=(\mL^n)_{>0},\\
\frac{\pa \mL}{\pa s} =& \{ \log \mP, \mL\}
,\nonumber\\
\frac{\pa \log \mP}{\pa t_n}=&\frac{\pa (\mL^n)_{\geq 0} }{\pa s}
- \{\log\mP, \B_n\}.\nonumber
\end{align}
\footnote{We understand that $\log \mP$ is formerly $\log k^N +
\frac{a_1}{k} + \frac{a_2}{k^2} + \ldots$.}where
$(\mathcal{A})_{S} = (\sum_{i} A_i k^i)_{S} = \sum_{i \in S} A_i
k^i $ specifies the part of the power series $\mathcal{A}$ to
extract, and $\{ \cdot, \cdot\}$ is the Poisson bracket
\[
\{ f, g\} = \frac{\pa f}{\pa k} \frac{\pa g}{\pa x} -  \frac{\pa
g}{\pa k} \frac{\pa f}{\pa x}.
\]
As usual, the $t_1$ flow says that \[ \frac{\pa \mL}{\pa t_1} =
\frac{\pa \mL}{\pa x}. \] In other words, the dependence on $t_1$
and $x$ appear in the combination $t_1+x$. When there are no
dependence on $s$, the first equation in \eqref{equation} is the
dmKP hierarchy defined by \cite{Kuper, CT}. We introduce an extra
parameter $s$ and the second equation in \eqref{equation}
determines the dependence of $\mL$ on $s$ via the auxiliary
polynomial $\mP$. The last equation determines the $t_n$-flow of
$\mP$.
% As a matter
%of fact, the last equation is a consequence of the first two (see
%below).
By standard argument, we have the following zero curvature
equation
\begin{align*}
\frac{\pa \B_m}{\pa t_n} -\frac{\pa \B_n}{\pa t_m} + \{ \B_m,
\B_n\} =0
\end{align*}
and its dual form
\begin{align}\label{zerocurvature}
\frac{\pa (\mL^m)_{\leq 0}}{\pa t_n} -\frac{\pa (\mL^n)_{\leq
0}}{\pa t_m} - \{ (\mL^m)_{\leq 0}, (\mL^n)_{\leq 0}\} =0.
\end{align}
This gives us the consistency between the $t_n$ flows of $\mL$. To
prove the consistency between the $t_n$ and $s$ flows of $\mL$, we
first establish the following.

\begin{proposition}\label{phi}
There exist a function $\phi(t)$ such that
\begin{align}\label{consistent}
\frac{\pa \phi}{\pa t_n} = (\mL^n)_0.
\end{align}
\end{proposition}
\begin{proof}
 We have to check that we can consistently solve for $\phi$.
Using the dual form of the zero curvature equation
\eqref{zerocurvature}, we have
\begin{align*}
\frac{\pa (\mL^m)_{ 0}}{\pa t_n} -\frac{\pa (\mL^n)_{0}}{\pa t_m}
- \left(\{ (\mL^m)_{\leq 0}, (\mL^n)_{\leq 0}\}\right)_0 =0.
\end{align*}
However, the last term contains powers $\leq -1$ of $k$. Hence
\eqref{consistent} is established. $\phi$ is not unique since we
do not specify its dependence on $s$.
\end{proof}
Now we check that the $s$ and $t_n$ flows of $\mL$ are consistent.
We have
\begin{align*}
&\frac{\pa}{\pa s} \{ \B_n , \mL\} -\frac{\pa}{\pa t_n} \{ \log
\mP, \mL \} \\
=& \{ \frac{\pa \B_n}{\pa s}, \mL\} + \{ \B_n, \{\log \mP, \mL\}\}
-\{ \frac{\pa \log \mP}{\pa t_n}, \mL \} - \{\log\mP,
\{ \B_n, \mL\}\}\\
=& \{ \frac{\pa\B_n}{\pa s} - \frac{\pa \log \mP}{\pa t_n} -
\{\log \mP, \B_n\}, \mL\} .
\end{align*}
Now from the third equation in \eqref{equation}, we have
\[
\frac{\pa\B_n}{\pa s} - \frac{\pa \log \mP}{\pa t_n} - \{\log \mP,
\B_n\} = - \frac{\pa (\mL^n)_0}{\pa s}.
\]
This is independent of $k$. On the other hand hand, from the
second equation in \eqref{equation}, since the right hand side
contains powers $\leq -1$ of $k$, we have
\begin{align*}
\frac{\pa (\mL)_0}{\pa s} =0.
\end{align*}
Together with $\frac{\pa (\mL^n)_0}{\pa t_m} = \frac{\pa
(\mL^m)_0}{\pa t_n}$ (from the proof of Proposition \ref{phi}), we
have
\begin{align*}
\frac{\pa (\mL^n)_0}{\pa x \pa s } = \frac{\pa (\mL^1)_0}{\pa t_n
\pa s} =0.
\end{align*}
Hence $\frac{\pa (\mL^n)_0}{\pa s}$ is independent of $x$, and the
$s$ and $t_n$ flows of $\mL$ are consistent.

For the consistency of the $t_n$ flows of $\mP$, we use the zero
curvature condition, the equation $\frac{\pa (\mL^n)_0}{\pa t_m} =
\frac{\pa (\mL^m)_0}{\pa t_n}$ ( from the proof of Proposition
\ref{phi}), and the fact $\frac{\pa (\mL^n)_0}{\pa s}$ is
independent of $k$ and $x$,
\begin{align*}
&\frac{\pa}{\pa t_m} \left(\frac{\pa (\mL^n)_{\geq 0} }{\pa s} -
\{\log\mP, \B_n\} \right)-\frac{\pa}{\pa t_n} \left(\frac{\pa
(\mL^m)_{\geq 0} }{\pa s} - \{\log\mP, \B_m\} \right)\\
=&\frac{\pa }{\pa s} \left( \frac{\pa (\mL^n)_{> 0} }{\pa t_m }-
\frac{\pa (\mL^m)_{> 0} }{\pa t_n }+\frac{\pa (\mL^n)_{ 0} }{\pa
t_m }- \frac{\pa (\mL^m)_{ 0} }{\pa t_n }\right)-\{ \log\mP,
\frac{\pa \B_n}{\pa t_m} -\frac{\pa \B_m}{\pa t_n}\}\\
&-\{ \frac{\pa (\mL^m)_{\geq 0} }{\pa s} - \{\log\mP, \B_m\} ,
\B_n\} + \{\frac{\pa (\mL^n)_{\geq 0} }{\pa s} - \{\log\mP,
\B_n\}, \B_m\}\\
=& \frac{\pa }{\pa s} \left( \frac{\pa \B_n }{\pa t_m }- \frac{\pa
\B_m }{\pa t_n } +\{\B_n, \B_m\}\right) -\{\log \mP, \frac{\pa
\B_n }{\pa t_m }- \frac{\pa \B_m }{\pa t_n } +\{\B_n, \B_m\}\}=0
\end{align*}
Hence the $t_n$ flows of $\mP$ are consistent.

\subsection{Dressing operator}
As in \cite{TT1}, we can establish the existence of a dressing
operator $\exp (\ad \varphi)$, where $\ad f (g) = \{f, g\}$.

\begin{proposition}\label{varphi}
There exists a function $\varphi = \sum_{n=0}^{\infty} \varphi_n
k^{-n}$, such that
\begin{align*}
\mL &= e^{\ad \varphi} k, \\
\nabla_{t_n, \varphi} \varphi &= - (\mL^n)_{\leq 0}, \hspace{1cm}
\nabla_{s, \varphi} \varphi = \log \mP - \log \mL^N,
\end{align*}
where $\nabla_{t, A} B = \sum_{n=0}^{\infty} \frac{(\ad
\varphi)^n}{(n+1)!} \frac{\pa A}{\pa t}$. If $\tilde{\varphi}$ is
another function satisfying the conditions above, then the
function $\psi= \sum_{n=0}^{\infty} \psi_n k^{-n}$ defined by
$e^{\ad \tilde{\varphi}} = e^{\ad \varphi} e^{\ad \psi}$ has
constant coefficients, i.e., $\psi_n$'s are independent of $t$.
\end{proposition}
\begin{proof}
We sketch the proof here. For details, we refer to Proposition
1.2.1 in \cite{TT1}. Standard argument shows that we can find
$\varphi_0$ such that
\begin{align*}
\mL &= e^{\ad \varphi_0} k. \end{align*} The first two equations
in \eqref{equation} imply that
\begin{align*}
\{ e^{-\ad \varphi_0} \left(\nabla_{t_n, \varphi_0} \varphi_0
+(\mL^n)_{\leq 0}\right), k\} =0,\\
\{ e^{-\ad \varphi_0} \left(\nabla_{s, \varphi_0} \varphi_0 -\log
\mP + \log \mL^N \right), k\} =0.
\end{align*}
Hence $A_n =e^{-\ad \varphi_0} \left(\nabla_{t_n, \varphi_0}
\varphi_0 +(\mL^n)_{\leq 0}\right)$ and $A_0= e^{-\ad \varphi_0}
\left(\nabla_{s, \varphi_0} \varphi_0 -\log \mP + \log \mL^N
\right)$ are independent of $x$. Using Lemma A.3 in the Appendix A
of \cite{TT1} and that $A_i$'s are independent of $x$, we have
$\frac{\pa A_n}{\pa t_m} = \frac{\pa A_m}{\pa t_n}$ and $\frac{\pa
A_n}{\pa s} = \frac{\pa A_0}{\pa t_n}$. Hence we can find a
function $\varphi'= \sum_{n=0}^{\infty} \varphi'_n k^{-n}$ such
that
\begin{align*}
\frac{\pa \varphi'}{\pa t_n} = -A_n,  \hspace{1cm} \frac{\pa
\varphi'}{\pa s} = -A_0.
\end{align*}
Moreover, $\varphi'$ can be chosen so that it is independent of
$x$. Now we define $\varphi$ by $e^{\ad \varphi} = e^{\ad
\varphi_0} e^{\ad \varphi'}$. Since $\varphi'$ is independent of
$x$, we have $e^{\ad \varphi'} k =k$ and $\nabla_{t_n, \varphi'}
\varphi' = \frac{\pa \varphi'}{\pa t_n}$, $\nabla_{s, \varphi'}
\varphi' = \frac{\pa \varphi'}{\pa s}$. So we have
\[
\mL = e^{\ad \varphi} k.
\]
Moreover, using Lemma A.2 in Appendix A of \cite{TT1}, we have
\begin{align*}
\nabla_{t_n, \varphi} \varphi &= \nabla_{t_n, \varphi_0} \varphi_0
+ e^{\ad \varphi_0} \nabla_{t_n, \varphi'} \varphi'  =
-(\mL^n)_{\leq 0},\\
\nabla_{s, \varphi} \varphi &=\nabla_{s, \varphi_0} \varphi_0 +
e^{\ad \varphi_0} \nabla_{s, \varphi'} \varphi'= \log \mP - \log
\mL^N.
\end{align*}
\end{proof}

\subsection{Orlov function and Darboux coordinates}
Using the dressing operator $\varphi$, we can construct the Orlov
function $\M$ by
\begin{align*}
\M =& e^{\ad \varphi} \left(\sum_{n=1}^{\infty} nk^{n-1} + x +
\frac{Ns}{k} \right)\\
=& \sum_{n=1}^{\infty} n\mL^{n-1} + x + \frac{Ns}{\mL} +
\sum_{n=1}^{\infty} v_n \mL^{-n-1},
\end{align*}
where $v_n$ are functions of $t$. $\M$ has the property that it
forms a canonical pair with $\mL$, namely $\{ \mL, \M\}=1$. Using
Proposition \ref{varphi} and Lemma A.1 in Appendix A of
\cite{TT1}, we also find
\begin{align}\label{equation2}
\frac{\pa \M}{\pa t_n} = \{ \B_n , \M\}, \hspace{1cm} \frac{\pa
\M}{\pa s} = \{ \log \mP, \M\}.
\end{align}

Now we want to express $B_n$'s and $\log\mP$ by $v_n$'s. From
definition, we can write the functions $\B_n$ and $\log \mP$ as
\begin{align*}
\B_n =& \mL^n -\frac{\pa \phi}{\pa t_n} -\sum_{m=1}^{\infty}
a_{n,m}(t)
\mL^{-m},\\
\log \mP =& \log \mL^N - \sum_{m=1}^{\infty} a_{0,m}(t) \mL^{-m},
\end{align*}
for some functions $a_{n,m}(t)$ of $t$. They  can be expressed in
terms of partial derivatives of $v_n$'s with respect to $t$:

\begin{proposition}\label{BandP}
\begin{align*}
\B_n =&\mL^n -\frac{\pa \phi}{\pa t_n} -\sum_{m=1}^{\infty}
\frac{1}{m}\frac{\pa v_m}{\pa t_n}
\mL^{-m},\\
\log \mP =& \log \mL^N - \sum_{m=1}^{\infty} \frac{1}{m} \frac{\pa
v_m}{\pa s} \mL^{-m}.
\end{align*}
\end{proposition}
\begin{proof}
First we have
\begin{align}\label{l1}
\frac{\pa \mL}{\pa t_n} =& \frac{\pa \B_n}{\pa k}\frac{\pa
\mL}{\pa
x} - \frac{\pa \B_n}{\pa x}\frac{\pa \mL}{\pa k}\\
=&\frac{\pa \B_n}{\pa \mL}\frac{\pa \mL}{\pa k}\frac{\pa \mL}{\pa
x} - \left(\frac{\pa \B_n}{\pa \mL}\frac{\pa \mL}{\pa x}+\frac{\pa
\B_n}{\pa x}\Bigr\vert_{\mL \fixed}\right)\frac{\pa \mL}{\pa k} \nonumber\\
=& -\frac{\pa \B_n}{\pa x}\Bigr\vert_{\mL \fixed}\frac{\pa
\mL}{\pa k}.\nonumber
\end{align}

The formulas
\[
\{\mL, \M\} =1 \hspace{1cm} \text{and} \hspace{1cm}\frac{\pa
\M}{\pa t_n} = \{\B_n, \M\}
\]
 satisfied by $\M$  give us
\begin{align*}
\frac{\pa \M}{\pa \mL}\frac{\pa \mL}{\pa t_n} + \frac{\pa \M}{\pa
t_n} \Bigr\vert_{\mL \fixed} &= \frac{\pa \B_n}{\pa \mL}\frac{\pa
\mL}{\pa k} \frac{\pa \M}{\pa x} - \left(\frac{\pa \B_n}{\pa
\mL}\frac{\pa \mL}{\pa x} + \frac{\pa \B_n}{\pa x}\Bigr\vert_{\mL
\fixed}\right) \frac{\pa \M}{\pa k}\\
&= \frac{\pa \B_n}{\pa \mL}-\frac{\pa \B_n}{\pa x}\Bigr\vert_{\mL
\fixed}\frac{\pa \M}{\pa \mL}\frac{\pa \mL}{\pa k}.
\end{align*}
In view of \eqref{l1}, we have
\begin{align*}
\frac{\pa \M}{\pa t_n} \Bigr\vert_{\mL \fixed}=\frac{\pa \B_n}{\pa
\mL}= n\mL^{n-1} +\sum_{m=1}^{\infty} m a_{n,m}\mL^{-m-1}.
\end{align*}
Comparing coefficients give us the assertion. The proof for the
formula for $\log \mP$ is the same, with $\log \mP$ replacing
$\B_n$, $s$ replacing $t_n$.
\end{proof}

 The proof of this proposition also gives us a characterization of the function $\M$ as
follows.
\begin{proposition}\label{orlov}
If $\M= \sum_{n=1}^{\infty} n\mL^{n-1} + x + \frac{Ns}{\mL} +
\sum_{n=1}^{\infty} v_n \mL^{-n-1}$ is a function such that
\begin{align*}
\{\mL, \M\}=1, \hspace{1cm} \frac{\pa \M}{\pa t_n} = \{\B_n ,
\M\}, \hspace{1cm} \frac{\pa \M}{\pa s} = \{ \log \mP, \M\},
\end{align*}
then there exists a dressing function $\varphi$ as in Proposition
\ref{varphi} and satisfies
\begin{align*}
\M = e^{\ad \varphi} \left(\sum_{n=1}^{\infty} nk^{n-1} + x +
\frac{Ns}{k} \right).
\end{align*}
In other words, $\M$ is an Orlov function of $\mL$.
\end{proposition}
\begin{proof}
We let $\varphi'$ be a dressing function given by Proposition
\ref{varphi} and define
\begin{align*}
\M' =& e^{\ad \varphi'} \left(\sum_{n=1}^{\infty} nt_nk^{n-1} + x
+
\frac{Ns}{k} \right)\\
=& \sum_{n=1}^{\infty} n\mL^{n-1} + x + \frac{Ns}{\mL} +
\sum_{n=1}^{\infty} v_n' \mL^{-n-1}.
\end{align*}
Then we have
\begin{align*}
\{\mL, \M'\} =1, \hspace{1cm}\frac{\pa \M'}{\pa t_n} = \{ \B_n ,
\M'\}, \hspace{1cm} \frac{\pa \M'}{\pa s} = \{ \log \mP, \M'\}.
\end{align*}
From the proof in the previous proposition, we see that these
conditions imply that
 \begin{align}\label{co}
\frac{\pa \M'}{\pa t_n}\Bigr\vert_{\mL \fixed} &= \frac{\pa
\B_n}{\pa \mL},
\hspace{2cm}\frac{\pa \M}{\pa t_n}\Bigr\vert_{\mL \fixed} = \frac{\pa \B_n}{\pa \mL},\\
\frac{\pa \M'}{\pa s}\Bigr\vert_{\mL \fixed} &= \frac{\pa \log
\mP}{\pa \mL}, \hspace{1.6cm}\frac{\pa \M}{\pa s}\Bigr\vert_{\mL
\fixed} = \frac{\pa \log \mP}{\pa \mL}.
\end{align}
From the explicit expressions of $\M$ and $\M'$, we have
\[
\M' -\M = \sum_{m=1}^{\infty} c_m \mL^{-m-1}.
\]
The equations in \eqref{co} imply that the $c_m$'s are constants.
We let $\varphi_0 = \sum_{m=1}^{\infty} \frac{c_m}{m}  k^{-m}$,
then
\begin{align*}
e^{\ad \varphi_0} \left(\sum_{n=1}^{\infty} nt_nk^{n-1} + x +
\frac{Ns}{k} \right)=\sum_{n=1}^{\infty} nt_nk^{n-1} + x +
\frac{Ns}{k}-\sum_{m=1}^{\infty} c_m k^{-m-1}.
\end{align*}
Hence if we define $\varphi$ by $e^{\ad \varphi} = e^{\ad
\varphi'} e^{\ad \varphi_0}$, then
\begin{align*}
e^{\ad \varphi} \left(\sum_{n=1}^{\infty} nt_nk^{n-1} + x +
\frac{Ns}{k} \right)=&e^{\ad \varphi'}\left(\sum_{n=1}^{\infty}
nt_nk^{n-1} + x + \frac{Ns}{k}-\sum_{m=1}^{\infty} c_m
k^{-m-1}\right) \\
=&\M' -\sum_{m=1}^{\infty} c_m \mL^{-m-1}=\M.
\end{align*}
Since $\varphi_0$ has constant coefficients, $\varphi$ satisfies
all the requirements in Proposition \ref{varphi}.
\end{proof}

Now we introduce the fundamental two form
\begin{align*}
\omega = dk \wedge dx + \sum_{n=1}^{\infty} d\B_n \wedge dt_n + d(
\log \mP - \frac{\pa \phi}{\pa s} ) \wedge ds.
\end{align*}
The exterior derivative $d$ is taken with respect to the
independent variables $k, x, s$ and $t_n, n=1,2, \ldots$. It
satisfies
\[
d\omega =0, \hspace{1cm} \text{and} \hspace{1cm} \omega \wedge
\omega =0.\] In fact, $(\mL, \M)$ is a pair of functions that play
the role of Darboux coordinates. Namely
\[
d\mL \wedge d\M = \omega.
\]
\begin{proposition}\label{omega}
The system of equations \eqref{equation} and \eqref{equation2},
together with $\{\mL, \M\}=1$  are equivalent to
\begin{align}\label{twoform}
d\mL \wedge d\M =dk \wedge dx + \sum_{n=1}^{\infty} d\B_n \wedge
dt_n + d( \log \mP - \frac{\pa \phi}{\pa s} ) \wedge ds.
\end{align}
\end{proposition}
\begin{proof}
We only show that \eqref{twoform} implies \eqref{equation} and
\eqref{equation2}, together with $\{\mL, \M\}=1$. The other
direction follows by tracing the argument backward. By looking at
the coefficients of $dk\wedge dx$, $dk\wedge dt_n$, $dx\wedge
dt_n$, $dk\wedge ds$, $dx \wedge ds$ and $dt_n \wedge ds$
respectively and using the property of $\phi$ given by Proposition
\ref{phi}, we find that \eqref{twoform} gives
\begin{align*}
\frac{\pa \mL}{\pa k}\frac{\pa \M}{\pa x} - \frac{\pa \mL}{\pa
x}\frac{\pa\M}{\pa k} &= \{\mL, \M\} =1,\\
\frac{\pa \mL}{\pa k}\frac{\pa \M}{\pa t_n} -\frac{\pa \M}{\pa
k}\frac{\pa \mL}{\pa t_n} &= \frac{\pa B_n}{\pa k},\hspace{1.2cm}
\frac{\pa \mL}{\pa x}\frac{\pa \M}{\pa t_n} -\frac{\pa \M}{\pa
x}\frac{\pa \mL}{\pa t_n} = \frac{\pa B_n}{\pa x},\\
\frac{\pa \mL}{\pa k}\frac{\pa \M}{\pa s} - \frac{\pa \M}{\pa
k}\frac{\pa\mL}{\pa s} &=\frac{\pa \log \mP}{\pa k},\hspace{1cm}
\frac{\pa \mL}{\pa x}\frac{\pa \M}{\pa s} - \frac{\pa \M}{\pa
x}\frac{\pa\mL}{\pa s} =\frac{\pa \log \mP}{\pa x}, \\
\frac{\pa \mL}{\pa s}\frac{\pa \M}{\pa t_n} -\frac{\pa \M}{\pa
s}\frac{\pa \mL}{\pa t_n} &=-\frac{\pa \log \mP}{\pa t_n} +
\frac{\pa^2 \phi}{\pa s \pa t_n}+ \frac{\pa \B_n}{\pa
s}=-\frac{\pa \log \mP}{\pa t_n} + \frac{\pa (\mL^n)_{\geq0}}{\pa
s}.
\end{align*}
The equations in the second line combine to give
\begin{align*}
\ma{-\frac{\pa \M}{\pa k}}{\frac{\pa \mL}{\pa k}}{-\frac{\pa
\M}{\pa x}}{\frac{\pa \mL}{\pa x}} \ve{\frac{\pa \mL}{\pa
t_n}}{\frac{\pa \M}{\pa t_n}}=\ve{\frac{\pa \B_n}{\pa
k}}{\frac{\pa B_n}{\pa x}},
\end{align*}
or since $\{\mL, \M\}=1$
\begin{align*}
 \ve{\frac{\pa \mL}{\pa
t_n}}{\frac{\pa \M}{\pa t_n}}=\ma{\frac{\pa \mL}{\pa
x}}{-\frac{\pa \mL}{\pa k}}{\frac{\pa \M}{\pa x}}{-\frac{\pa
\M}{\pa k}}\ve{\frac{\pa \B_n}{\pa k}}{\frac{\pa B_n}{\pa x}}
\end{align*}
i.e.
\begin{align*}
\frac{\pa \mL}{\pa t_n} = \{\B_n, \mL\}, \hspace{1cm} \frac{\pa
\M}{\pa t_n} = \{\B_n, \M\}.
\end{align*}
Similarly, the equations in the third line combine to give
\begin{align*}
\frac{\pa \mL}{\pa s} = \{\log\mP, \mL\}, \hspace{1cm} \frac{\pa
\M}{\pa s} =\{\log\mP, \M\}.
\end{align*}
Using the previous result, the last equation gives
\begin{align*}
-\frac{\pa \log \mP}{\pa t_n} + \frac{\pa (\mL^n)_{\geq0}}{\pa
s}=& \left( \frac{\pa \log\mP}{\pa k}\frac{\pa \mL}{\pa x} -
\frac{\pa \log\mP}{\pa x}\frac{\pa \mL}{\pa k}\right)
\left(\frac{\pa \B_n}{\pa k}\frac{\pa \M}{\pa x}-\frac{\pa
\B_n}{\pa x}\frac{\pa \M}{\pa
k}\right)\\
&-\left( \frac{\pa \log\mP}{\pa k}\frac{\pa \M}{\pa x} - \frac{\pa
\log\mP}{\pa x}\frac{\pa \M}{\pa k}\right) \left(\frac{\pa
\B_n}{\pa k}\frac{\pa \mL}{\pa x}-\frac{\pa \B_n}{\pa x}\frac{\pa
\mL}{\pa k}\right)\\
=&\{\log \mP, \B_n\}.
\end{align*}
The coefficients of $dt_n \wedge dt_m$ gives the zero curvature
condition.
\end{proof}

\subsection{$S$ function and tau function}
Proposition \ref{omega} implies that we can find a function $S$
such that
\begin{align*}
dS = \M d\mL + kdx + \sum_{n=1}^{\infty} \B_n dt_n + (\log
\mP-\frac{\pa \phi}{\pa s}) ds.
\end{align*}
In other words,
\begin{align*}
\frac{\pa S}{\pa \mL} = \M, \hspace{0.5cm} \frac{\pa S}{\pa
x}\Bigr\vert_{\mL \fixed}=k ,\hspace{0.5cm}\frac{\pa S}{\pa
t_n}\Bigr\vert_{\mL \fixed}=\B_n ,\hspace{0.5cm}\frac{\pa S}{\pa
s}\Bigr\vert_{\mL \fixed}=\log\mP -\frac{\pa \phi}{\pa s}.
\end{align*}
The second equation is just a special case of the third when $n=1$
since $\B_1=k$. From the explicit representation of the function
$\B_n$'s and $\log\mP$ given by Proposition \ref{BandP}, we have
the following explicit expression for $S$.

\begin{proposition}\label{Sfunction}
\[
S = \sum_{n=1}^{\infty} t_n \mL^n + x \mL + s\log \mL^N +
\sum_{n=1}^{\infty} -\frac{v_n}{n} \mL^{-n} -\phi.
\]
\end{proposition}

We also have the following characterization of the partial
derivatives of $v_n$'s in terms of residues.
\begin{proposition}\label{duality}
\begin{align*}
 \frac{\pa v_n}{\pa t_m} = \Res \mL^n d_k\B_m, \hspace{2cm}
 \frac{\pa v_n}{\pa s} = \Res \mL^nd_k\log\mP.
\end{align*}
Here the differential $d_k$ is taken with respect to $k$ and $\Res
A dk$ means the coefficient of $k^{-1}$ of $A$.
\end{proposition}
\begin{proof}
We give a proof which follows the same line as Proposition 4 in
\cite{TT2}. We only show the second equality here. We have
\begin{align*}
\frac{\pa\M}{\pa s} - \frac{\pa \M}{\pa \mL} \frac{\pa \mL}{\pa s}
=\frac{\pa \M}{\pa s} \Bigr\vert_{\mL \fixed} = \frac{N}{\mL} +
\sum_{n=1}^{\infty} \frac{\pa v_n}{\pa s} \mL^{-n-1}.
\end{align*}
Using $\Res \mL^n d_k\mL = \delta_{n,-1}$, we have
\begin{align*}
\frac{\pa v_n}{\pa s} &= \Res \mL^n \left(\frac{\pa\M}{\pa s} -
\frac{\pa \M}{\pa \mL}
\frac{\pa \mL}{\pa s}  \right) d_k\mL.\\
\end{align*}
The expression in the bracket
\begin{align*}
& \left(\frac{\pa\M}{\pa s} - \frac{\pa \M}{\pa \mL}
\frac{\pa \mL}{\pa s}\right) d_k\mL\\
 =& \Biggl(\{\log \mP, \M\} \frac{\pa
\mL}{\pa k}-\{\log\mP, \mL\}
\frac{\pa \M}{\pa \mL}\frac{\pa \mL}{\pa k}\Biggr)dk\\
=&\Biggl(\left(\frac{\pa \log\mP}{\pa k} \frac{\pa \M}{\pa
x}-\frac{\pa \log\mP}{\pa x} \frac{\pa \M}{\pa k}\right) \frac{\pa
\mL}{\pa k}-\left(\frac{\pa \log\mP}{\pa k} \frac{\pa \mL}{\pa
x}-\frac{\pa \log\mP}{\pa x} \frac{\pa \mL}{\pa k}\right)
\frac{\pa \M}{\pa k}\Biggr)dk \\
=&\frac{\pa \log \mP}{\pa k}dk = d_k \log\mP.
\end{align*}
This implies the assertion.
\end{proof}

 As a consequence of  this proposition, we can show the existence
 of a tau function for our dcmKP hierarchy. First
 we have the following properties of taking residues,
 \begin{align*}
 \Res Ad_kB =& -\Res Bd_kA,\\
\Res A d_kB =& \Res A_{>0} d_k B_{<0} + \Res A_{<0} d_k B_{>0}
\end{align*}
When $m, n \geq 1$, we have $\Res \mL^m d_k \mL^n = n\Res
\mL^{m+n-1} d_k\mL =0$. Hence
 \begin{align*}
\Res (\mL^m)_{>0}d_k(\mL^n)_{<0}= - \Res
 (\mL^m)_{<0}d_k(\mL^n)_{>0}.
 \end{align*}
 It follows that
 \begin{align}\label{con1}
\frac{\pa v_m}{\pa t_n} =&\Res \mL^m d_k\B_n= \Res (\mL^m)_{<0}
d_k(\mL_n)_{>0} = -\Res (\mL^m)_{>0} d_k(\mL^n)_{<0}\\
=&\Res (\mL^n)_{<0} d_k(\mL^m)_{>0} =\frac{\pa v_n}{\pa t_m}.
\nonumber
 \end{align}
On the other hand, since
\begin{align*}
\frac{\pa }{\pa t_m} \Res \mL^n d_k\log\mP =\frac{\pa^2 v_n}{\pa
s\pa t_m}=\frac{\pa^2 v_m}{\pa s \pa t_n} = \frac{\pa}{\pa t_n}
\Res \mL^m d_k\log \mP,
\end{align*}
there exists a function $\Phi(t)$ such that
\begin{align}\label{con2}
\frac{\pa \Phi}{\pa t_n} = \Res \mL^n d_k\log\mP=\frac{\pa
v_n}{\pa s}.
\end{align}
 $\Phi$
is only determined up to a function of $s$. Finally from
\eqref{con1} and \eqref{con2}, we have
 \begin{proposition}
There exist a function $\tau$ of $t$, called the tau function of
our dcmKP hierarchy, such that $\log \tau$ is the generating
function for $v_n$'s and $\Phi$, i.e.
\[
\frac{\pa \log \tau}{\pa t_n} = v_n, \hspace{2cm} \frac{\pa \log
\tau}{\pa s} = \Phi.
\]
\end{proposition}

Since $\Phi$ is only determined up to a function of $s$, $\tau$ is
also only determined up to a function of $s$ \footnote{This degree
of freedom can also be viewed as due to we do not specify the $s$
flow of $\mP$.}.

\begin{remark}
In the special case when $\mP = k$, we have
\[
\Res \mL^n d_k\log k = (\mL^n)_0.
\]
Hence the function $\Phi$ coincides with the function $\phi$ we
introduce in Proposition \ref{phi}.

In this case, the formulas in Proposition \ref{BandP} can be
rewritten in terms of the tau function as
\begin{align}\label{special}
\B_n &= \mL^n -\sum_{m=1}^{\infty} \frac{1}{m}\frac{\pa^2 \log
\tau}{\pa
t_m\pa t_n} \mL^{-m} - \frac{\pa^2\log \tau}{\pa s \pa t_n},\\
\log k &=\log \mL - \sum_{m=1}^{\infty} \frac{1}{m}\frac{\pa^2
\log \tau}{\pa t_m \pa s}\mL^{-m}. \nonumber
\end{align}
Hence the coefficients of $\mL, \M, \B_n$ can be expressed as
differential polynomials of derivatives of $\log \tau.$ We shall
see below that this special case play a particular role of
bridging the transition from dispersionless Toda (dToda) hierarchy
to dispersionless KP (dKP) hierarchy. This is the motivation for
our present work.
\end{remark}
\section{Relations with dispersionless Toda hierarchy}
There are a few ways to formulate dispersionless Toda hierarchy,
All of them are connected by a Miura type transformation. We first
quickly review the set up following \cite{TT1}. For details, we
refer to \cite{TT1} and references therein.

dToda is a system of differential equations with two sets of
independent variables $(t_1, t_2, \dots)$,$(t_{-1}, t_{-2},
\ldots)$ and an independent variable $s$. The fundamental
quantities are two formal power series
\begin{align*}
\mL(p)=& p + \sum_{n=0}^{\infty} u_{n+1}(t) p^{-n}\\
\tilde{\mL}^{-1}(p) =& \tilde{u}_0(t)p^{-1} + \sum_{n=0}^{\infty}
\tilde{u}_{n+1}(t) p^n.
\end{align*}
Here $t$ denotes collectively all the independent variables. The
Lax representation is
\begin{align*}
\frac{\pa \mL}{\pa t_n} =\{ (\mL^n)_{\geq 0}, \mL\}_T,
\hspace{2cm}
\frac{\pa \mL}{\pa t_{-n}} = \{(\tilde{\mL}^{-n})_{<0}, \mL\}_T,\\
\frac{\pa \tilde{\mL}}{\pa t_n} =\{ (\mL^n)_{\geq 0},
\tilde{\mL}\}_T, \hspace{2cm} \frac{\pa \tilde{\mL}}{\pa t_{-n}} =
\{(\tilde{\mL}^{-n})_{<0}, \tilde{\mL}\}_T.
\end{align*}
Here $\{ \cdot, \cdot\}_T$ is the Poisson bracket for dToda
hierarchy
\[
\{f, g\}_T =p \frac{\pa f}{\pa p} \frac{\pa g}{\pa s}-p\frac{\pa
f}{\pa s} \frac{\pa g}{\pa p}.
\]
In order to see the relation between the dToda hierarchy and our
dcmKP hierarchy, we use the fact that (for details see \cite{TT1}
) there exist a tau function $\tau_{dToda}$ such that in terms of
this tau function,
\begin{align}\label{tau}
(\mL^n)_{\geq 0} & = \mL^n - \sum_{m=1}^{\infty}
\frac{1}{m}\frac{\pa^2 \log \tau_{\dToda}}{\pa
t_m\pa t_n} \mL^{-m}, \\
(\mL^n)_0 &=\frac{\pa^2 \log \tau_{\dToda}}{\pa s \pa t_n},\nonumber\\
 \log p
&=\log \mL - \sum_{m=1}^{\infty} \frac{1}{m}\frac{\pa^2 \log
\tau_{\dToda}}{\pa t_m \pa s}\mL^{-m}\nonumber.
\end{align}
Comparing with equations \eqref{special}, it is natural to see
that the following proposition should hold.
\begin{proposition}
If $(\mL, \tilde{\mL})$ is a solution to dToda and $\frac{\pa
(\mL)_0}{\pa s} =0$, then $(\mL, \mP=k)$ is also a solution to
dcmKP hierarchy if we replace $p$ by $k$, $t_1$ by $t_1 +x$ and
regarding $t_{-n}$'s as parameters. In this case, the tau function
for the dToda hierarchy $\tau_{\dToda}$ is also the tau function
for the corresponding dcmKP hierarchy.
\end{proposition}
\begin{proof}

Replacing $p$ by $k$ and $t_1$ by $t_1+x$, the case $n=1$ of the
first equation in the dToda hierarchy gives us
\begin{align*}
\frac{\pa \mL}{\pa x}=\frac{\pa \mL}{\pa t_1} = k\frac{\pa
(\mL)_{\geq 0}}{\pa k}\frac{\pa \mL}{\pa s} - k\frac{\pa
(\mL)_{\geq 0}}{\pa s}\frac{\pa \mL}{\pa k}.
\end{align*}
Now $(\mL)_{\geq 0} = k+u_1$. Since we assume that $u_1$ is
independent of $s$, $\frac{\pa (\mL)_{\geq 0}}{\pa s}$ is
identically $0$. Hence
\begin{align*}
\frac{\pa \mL}{\pa x} = k\frac{\pa \mL}{\pa s},
\end{align*}
or equivalently,
\begin{align*}
\frac{\pa \mL}{\pa s} = \{\log k, \mL\}=\frac{1}{k}\frac{\pa
\mL}{\pa x},
\end{align*}
which is the second equation in our dcmKP hierarchy
\eqref{equation} with $\mP=k$.

From this equation, we also have
\begin{align*}
\frac{\pa \mL^n}{\pa x} = k\frac{\pa \mL^n}{\pa s}.
\end{align*}
Comparing powers of $k$, we have
\begin{align*}
\frac{\pa (\mL^n)_{>0}}{\pa x} = k\frac{\pa (\mL^n)_{\geq 0}}{\pa
s},
\end{align*}
or equivalently,

 \begin{align*}
\frac{\pa (\mL^n)_{\geq 0}}{\pa s}=\frac{1}{k}\frac{\pa
(\mL^n)_{>0}}{\pa x}=\{\log k, (\mL^n)_{>0}\},
 \end{align*}
 which is the third equation in our dcmKP hierarchy
 \eqref{equation}.
 Finally using the other equations in the dToda hierarchy, we have
 \begin{align*}
\frac{\pa \mL}{\pa t_n} &= k\frac{\pa (\mL^n)_{\geq 0}}{\pa
k}\frac{\pa \mL}{\pa s}- k\frac{\pa (\mL^n)_{\geq 0}}{\pa
s}\frac{\pa \mL}{\pa k} =\frac{\pa (\mL^n)_{\geq 0}}{\pa
k}\frac{\pa \mL}{\pa x}- \frac{\pa (\mL^n)_{> 0}}{\pa x}\frac{\pa
\mL}{\pa
k}\\
&=\frac{\pa (\mL^n)_{> 0}}{\pa k}\frac{\pa \mL}{\pa x}- \frac{\pa
(\mL^n)_{> 0}}{\pa x}\frac{\pa \mL}{\pa k}=\{(\mL^n)_{>0}, \mL\},
 \end{align*}
which is the first equation in our dcmKP hierarchy
\eqref{equation}.
\end{proof}
\begin{remark}
In the proof of this proposition, we also see that in the special
case when $\mP =k$, the third equation in \eqref{equation} is a
consequence of the second equation.
\end{remark}

\section{Miura map between dmKP and dKP hierarchies}

We establish that if $(\mL, \mP)$ is a solution of our dcmKP
hierarchy, then via a Miura transform, it will give a solution of
the dmKP hierarchy defined by Takebe \cite{T1}. The fundamental
quantity in Takebe's definition is the formal series
\begin{align*}
\tilde{\mL} = k +\sum_{n=1}^{\infty} \tilde{u}_{n+1}(t) k^{-n},
\end{align*}
and an auxiliary polynomial \[ \tilde{\mP} = k^N +
\tilde{p}_{N-1}(t) k^{N-1} + \ldots+ \tilde{p}_0(t).
\]
Here $t=(x, s, t_1 , t_2, \dots)$ are independent variables. The
Lax representation is
\begin{align}\label{equation4}
\frac{\pa\tilde{\mL}}{\pa t_n} =& \{(\tilde{\mL}^n)_{\geq 0},
\tilde{\mL}\}, \hspace{1cm} \frac{\pa \tilde{\mL}}{\pa s} = \{
\log \tilde{\mP}, \tilde{\mL}\}
,\\
\frac{\pa \log \tilde{\mP}}{\pa t_n} =&\frac{\pa
(\tilde{\mL}^n)_{\geq 0} }{\pa s} -  \{\log\tilde{\mP},
(\tilde{\mL}^n)_{\geq 0}\}. \nonumber
\end{align}

Let $(\mL, \mP)$ be a solution of our dcmKP hierarchy
\eqref{equation} and $\phi$ be the function defined by Proposition
\ref{phi}. The Miura transform of $(\mL, \mP)$ is given by
\[
\tilde{\mL} = e^{\ad \phi} \mL, \hspace{1cm} \tilde{\mP} = e^{\ad
\phi} \mP.
\]
Since $\phi$ is independent of $k$, we find
\begin{align*}
e^{\ad \phi} \mL = e^{\ad \phi}k + \sum_{n=0}^{\infty} u_{n+1}(t)
(e^{\ad \phi} k)^{-n},
\end{align*}
and
\[
e^{\ad \phi} k = k - \frac{\pa \phi}{\pa x} = k-u_1.
\]
Hence
\begin{align*}
\tilde{\mL} = k-u_1(t) + \sum_{n=0}^{\infty} u_{n+1}(t)
(k-u_1(t))^{-n} = k + \sum_{n=1}^{\infty} \tilde{u}_{n+1}(t)
k^{-n}
\end{align*}
does not has term in $k^0$.

\begin{proposition}\label{Miura}
Let $(\mL, \mP)$ be a solution of the dcmKP hierarchy
\eqref{equation}, then $(\tilde{\mL}, \tilde{\mP})$ defined by the
Miura transform is a solution of the dmKP hierarchy defined by
Takebe \cite{T1}.
\end{proposition}
\begin{proof}
We prove that $(\tilde{\mL}, \tilde{\mP})$ satisfies the dmKP
hierarchy \eqref{equation4} defined by Takebe. Using the formulas
in Appendix A of \cite{TT1}, we have
\begin{align*}
\frac{\pa \tilde{\mL}}{\pa t_n} =& e^{\ad \phi} \frac{\pa\mL}{\pa
t_n} +
\{ \nabla_{t_n, \phi} \phi, \tilde{\mL}\}\\
=& \{e^{\ad \phi} (\mL^n)_{>0}, e^{\ad \phi} \mL\} +\{ \frac{\pa \phi}{\pa t_n}, \tilde{\mL}\}\\
\end{align*}
Now if we write
\[
\mL^n = \sum_{m=-\infty}^{n} v_{n, m}(t) k^{m}
\]
we have
\begin{align*}
\tilde{\mL}^n =\sum_{m=-\infty}^{n} v_{n, m}(t) (k-u_1)^{m}
\end{align*}
Hence
\begin{align}\label{zero}
(\tilde{\mL}^n)_{\geq 0} = \sum_{m=0}^{n} v_{n,m} (k-u_1)^{m} =
\sum_{m=1}^{\infty} v_{n,m} (k-u_1)^{m} +(\mL^n)_0 = e^{\ad \phi}
(\mL^n)_{>0} + \frac{\pa \phi}{\pa t_n}.
\end{align}
Hence we have established the first equation in \eqref{equation4}.
 Similarly, we have
\begin{align*}
\frac{\pa \tilde{\mL}}{\pa s} =& e^{\ad \phi} \frac{\pa\mL}{\pa s}
+
\{ \nabla_{s, \phi} \phi, \tilde{\mL}\}\\
=& \{e^{\ad \phi} \log \mP, e^{\ad \phi} \mL\} +\{ \frac{\pa
\phi}{\pa s}, \tilde{\mL}\}.
\end{align*}
Since $\frac{\pa \phi}{\pa s} $ is independent of $k$ and $x$, we
have established the second equation in \eqref{equation}. Finally,
\begin{align*}
\frac{\pa \log\tilde{\mP}}{\pa t_n} =& e^{\ad \phi} \frac{\pa \log
\mP}{\pa t_n} +
\{ \nabla_{t_n, \phi} \phi, \log \tilde{\mP}\}\\
=& e^{\ad \phi} \left(\frac{ \pa (\mL^n)_{\geq 0} }{\pa s}-
\{\log\mP, (\mL^n)_{>0}\} \right)+\{ (\mL^n)_0,
\log\tilde{\mP}\}\\
=&\frac{ \pa (\tilde{\mL}^n)_{\geq 0} }{\pa s}- \{\log\tilde{\mP},
e^{\ad \phi}(\mL^n)_{>0}\} - \{\log \tilde{\mP}, (\mL^n)_0\}\\
=&\frac{ \pa (\tilde{\mL}^n)_{\geq 0} }{\pa s}- \{\log\tilde{\mP},
(\tilde{\mL}^n)_{\geq0}\} .
\end{align*}
which is the third equation in \eqref{equation4}.
\end{proof}
Notice that if we regard $s$ as a parameter, then $\tilde{\mL}$ is
a solution of the dKP hierarchy.

\begin{remark}
To go in the opposite direction, we have
\begin{align}\label{Miura}
\mL = e^{-\ad \phi} \tilde{\mL}.
\end{align}
Express in terms of $\tilde{\mL}$, we have from \eqref{zero},
\[
\frac{\pa \phi}{\pa t_n} =(\mL^n)_0=(\tilde{\mL}^n)_{\geq
0}\Bigr\vert_{k=u_1}=(\tilde{\mL}^n)_{\geq
0}\Bigr\vert_{k=\frac{\pa \phi}{\pa x}}.
\]
This is precisely the condition that $\phi$ must satisfy in order
that \eqref{Miura} transforms a dKP solution to a dmKP solution
which is proved by Chang and Tu in \cite{CT}.

\end{remark}

%\begin{remark}
%If $(\mL, \tilde{\mL})$ is a solution of the dToda hierarchy,
%there exists a function $\phi(t)$ such that
%\begin{align*}
%\frac{\pa \phi}{\pa t_n} = (\mL^n)_0 , \hspace{1cm} \frac{\pa
%\phi}{\pa \s_n}=-(\tilde{\mL}^{-n})_0, \hspace{1cm} \frac{\pa
%\phi}{\pa s} =\log r.
%\end{align*}
%With this $\phi$, we can make a Miura-type transformation (Lemma
%2.1.3 in \cite{TT1}) and define $\mL' = e^{-\ad \phi}\mL$,
%$\tilde{\mL}'=e^{-\ad \phi}\tilde{\mL}$. Then they are of the form
%\begin{align*}
%\mL'(p)=& rp + \sum_{n=0}^{\infty} u'_{n+1}(t) p^{-n}\\
%\tilde{\mL'}^{-1}(p) =& p^{-1} + \sum_{n=0}^{\infty}
%\tilde{u}'_{n+1}(t) p^n,
%\end{align*}
%and satisfies the Lax equations
%\begin{align*}
%\frac{\pa \mL'}{\pa t_n} =\{ (\mL^n)_{> 0}, \mL\}_T, \hspace{2cm}
%\frac{\pa \mL'}{\pa \s_n} = \{(\tilde{\mL})_{\leq0}, \mL\}_T,\\
%\frac{\pa \tilde{\mL'}}{\pa t_n} =\{ (\mL^n)_{> 0},
%\tilde{\mL}\}_T, \hspace{2cm} \frac{\pa \tilde{\mL'}}{\pa \s_n} =
%\{(\tilde{\mL})_{\leq0}, \tilde{\mL}\}_T.
%\end{align*}
%\end{remark}

\begin{remark}
Notice that if $\varphi$ is the dressing operator for $\mL$, then
the function $\tilde{\varphi}$ defined by $e^{\ad \tilde{\varphi}}
= e^{\ad \phi} e^{\ad \varphi}$ is the dressing operator for
$\tilde{\mL}$. If we denote the Orlov function for $\tilde{\mL}$
by $\tilde{\M}$, then by definition (see \cite{T1})
\begin{align*}
\tilde{\M} =&e^{\ad \tilde{\varphi}}\left(\sum_{n=1}^{\infty} nt_n
k^{n-1} + x + \frac{Ns}{k}\right) = e^{\ad \phi} e^{\ad
\varphi}\left(\sum_{n=1}^{\infty} nt_n
k^{n-1} + x + \frac{Ns}{k}\right)\\
=&e^{\ad \phi} \left( \sum_{n=1}^{\infty} nt_n \mL^{n-1} + x+
\frac{Ns}{\mL} + \sum_{n=1}^{\infty} v_n \mL^{-n-1}\right)\\
=&\sum_{n=1}^{\infty} nt_n \tilde{\mL}^{n-1} + x+
\frac{Ns}{\tilde{\mL}} + \sum_{n=1}^{\infty} v_n
\tilde{\mL}^{-n-1}.
\end{align*}
Notice that the functions $v_n$'s are the same in both
hierarchies. In particular, the tau function for a solution of our
dcmKP hierarchy is also the tau function for the corresponding
dmKP hierarchy obtained via a Miura transform.
 \end{remark}

\section{Twistor construction}
As in \cite{TT1} and \cite{T1}, we show that a solution of dcmKP
can be obtained from twistor construction (or Riemann Hilbert
 type construction).

A twistor data for a solution $(\mL, \mP)$ of dcmKP hierarchy is a
pair of functions $(f, g)$ in variables $(k, x)$ such that $\{f,
g\} =1$ and
\[
f(\mL, \M) _{\leq 0} =0, \hspace{1.5cm} g(\mL, \M)_{<0}=0.
\]

\begin{proposition}\label{twistor}
Let $(f,g)$ be a pair of functions in $(k, x)$ given such that
$\{f,g\}=1$. Let
\begin{align*}
\mL &= k + \sum_{n=0}^{\infty}
u_{n+1}k^{-n},\\
 \M &= \sum_{n=1}^{\infty} nt_{n} \mL^{n-1} + x +
\frac{Ns}{\mL} + \sum_{n=1}^{\infty} v_n \mL^{-n-1},\\
\mP &= k^N + p_{N-1}(t) k^{N-1} + \ldots +p_0(t)
\end{align*} be
formal power series in $k$ with coefficients depending on $t$. If
\begin{align}\label{identity8}
f(\mL, \M)_{\leq 0} =0, \hspace{1.5cm}& g(\mL, \M)_{<0}=0,\\
\left( \{ \log \mP, f(\mL, \M)\}\right)_{\leq 0} =0,\hspace{1cm}&
\left(\{ \log \mP, g(\mL, \M)\}\right)_{< -1} =0 \nonumber,
\end{align}
Then $(\mL, \mP)$ is a solution to our dcmKP hierarchy with $\M$
the corresponding Orlov function.
\end{proposition}
\begin{proof}
For the case when $\mL$ is independent of $s$, the first two
conditions are those given by Chang and Tu \cite{CT} in order that
$\mL, \M$ satisfies $\{\mL, \M\}=1$, the first equations in
\eqref{equation} and the first equations in \eqref{equation2} of
the dcmKP hierarchy. The other two conditions (which are modified
by those given by Takebe in \cite{T1}\footnote{Notice that our $f$
and $g$ do not depend on $s$, though in general we can let them
have one more degree of freedom in $s$.}) are such that the other
equations in \eqref{equation} and \eqref{equation2} are satisfied.
For completeness, we give the full proof here.

We let
\begin{align}\label{def}
\wh{\mL}= f(\mL, \M) \hspace{1cm} \text{and} \hspace{1cm} \wh{\M}
= g(\mL, \M).
\end{align}
Then the condition of the proposition says that
\begin{align*}
\wh{\mL}_{\leq 0} =0, \hspace{0.5cm}\wh{\M}_{<0}=0, \hspace{0.5cm}
\{ \log \mP, \wh{\mL}\}_{\leq 0} =0,\hspace{0.5cm}& \{ \log \mP,
\wh{\M}\}_{< -1} =0,
\end{align*}
Taking derivative of the equations in \eqref{def} with respect to
$k$ and $x$, we have the following system:
\begin{align}\label{identity1}
\ma{\frac{\pa f}{\pa \mL}}{\frac{\pa f}{\pa \M}}{\frac{\pa g}{\pa
\mL}}{\frac{\pa g}{\pa \M}} \ma{\frac{\pa \mL}{\pa k}}{\frac{\pa
\mL}{\pa x}}{\frac{\pa \M}{\pa k}}{\frac{\pa \M}{\pa
x}}=\ma{\frac{\pa \wh{\mL}}{\pa k}}{\frac{\pa \wh{\mL}}{\pa
x}}{\frac{\pa \wh{\M}}{\pa k}}{\frac{\pa \wh{\M}}{\pa x}}
\end{align}
Taking determinant of both sides, since $\{f,g\}=1$, we have
\[
\{\mL, \M\} = \{\wh{\mL}, \wh{\M}\}.
\]
Now the left hand side
\begin{align*}
\frac{\pa \mL}{\pa k}\frac{\pa \M}{\pa x}- \frac{\pa \mL}{\pa
x}\frac{\pa \M}{\pa k} =&\frac{\pa \mL}{\pa k}\left(\frac{\pa
\M}{\pa \mL}\frac{\pa \mL}{\pa x}+ \frac{\pa \M}{\pa
x}\Bigr\vert_{\mL \fixed}\right)- \frac{\pa \mL}{\pa x}\frac{\pa
\M}{\pa \mL}\frac{\pa \mL}{\pa k}\\
=&\frac{\pa \mL}{\pa k} \left(1+ \sum_{m=1}^{\infty} \frac{\pa
v_m}{\pa x} \mL^{-m-1}\right)\\
 =&1+ (\text{powers $<0$ of $k$}).
\end{align*}
While the right hand side
\[
\frac{\pa \wh{\mL}}{\pa k}\frac{\pa \wh{\M}}{\pa x}- \frac{\pa
\wh{\mL}}{\pa x}\frac{\pa \wh{\M}}{\pa k}= (\text{powers $\geq 0$
of $k$}).
\]
Comparing powers of both sides, we have
\begin{align}\label{identity2}
\{\mL, \M\} &= \{\wh{\mL}, \wh{\M}\}=1.
%\hspace{1cm} \frac{\pa
%\mL}{\pa k} \left(1+ \sum_{m=1}^{\infty} \frac{\pa v_m}{\pa x}
%\mL^{-m-1}\right)=1.
\end{align}
Next, taking derivative of \eqref{def} with respect to $t_n$, we
have
\begin{align*}
\ma{\frac{\pa f}{\pa \mL}}{\frac{\pa f}{\pa \M}}{\frac{\pa g}{\pa
\mL}}{\frac{\pa g}{\pa \M}}\ve{\frac{\pa \mL}{\pa t_n}}{\frac{\pa
\M}{\pa t_n}} =\ve{\frac{\pa \wh{\mL}}{\pa t_n}}{\frac{\pa
\wh{\M}}{\pa t_n}}.
\end{align*}
Using \eqref{identity1} and \eqref{identity2}, this is equivalent
to
\begin{align*}
\ma{\frac{\pa \M}{\pa x}}{-\frac{\pa \mL}{\pa x}}{-\frac{\pa
\M}{\pa k}}{\frac{\pa \mL}{\pa k}}\ve{\frac{\pa \mL}{\pa
t_n}}{\frac{\pa \M}{\pa t_n}} =\ma{\frac{\pa \wh{\M}}{\pa
x}}{-\frac{\pa \wh{\mL}}{\pa x}}{-\frac{\pa \wh{\M}}{\pa
k}}{\frac{\pa \wh{\mL}}{\pa k}}\ve{\frac{\pa \wh{\mL}}{\pa
t_n}}{\frac{\pa \wh{\M}}{\pa t_n}}.
\end{align*}
Hence we have
\begin{align*}
\frac{\pa \M}{\pa x} \frac{\pa \mL}{\pa t_n} -\frac{\pa\mL}{\pa
x}\frac{\pa \M}{\pa t_n} =&\frac{\pa \wh{\M}}{\pa x} \frac{\pa
\wh{\mL}}{\pa t_n} -\frac{\pa\wh{\mL}}{\pa x}\frac{\pa
\wh{\M}}{\pa t_n}=(\text{powers $>0$ of $k$})\\
-\frac{\pa \M}{\pa k} \frac{\pa \mL}{\pa t_n} +\frac{\pa\mL}{\pa
k}\frac{\pa \M}{\pa t_n} =&-\frac{\pa \wh{\M}}{\pa k} \frac{\pa
\wh{\mL}}{\pa t_n} +\frac{\pa\wh{\mL}}{\pa k}\frac{\pa
\wh{\M}}{\pa t_n}=(\text{powers $\geq 0$ of $k$}).
\end{align*}
We can rewrite the left hand sides of these equations as
\begin{align*}
&\left(\frac{\pa \M}{\pa \mL}\frac{\pa \mL}{\pa x} + \frac{\pa
\M}{\pa x} \Bigr\vert_{\mL \fixed}\right) \frac{\pa \mL}{\pa
t_n}-\frac{\pa\mL}{\pa x}\left(\frac{\pa \M}{\pa \mL}\frac{\pa
\mL}{\pa t_n}+ \frac{\pa \M}{\pa t_n}\Bigr\vert_{\mL
\fixed}\right)\\
=& \left( 1+ \sum_{m=1}^{\infty}\frac{\pa v_m}{\pa
x}\mL^{-m-1}\right) \frac{\pa \mL}{\pa t_n} - \frac{\pa \mL}{\pa
x} \left( n\mL^{n-1} + \sum_{m=1}^{\infty} \frac{\pa v_m}{\pa t_n}
\mL^{-m-1}\right)\\
=& - \frac{\pa (\mL^n)_{>0}}{\pa x} + (\text{powers $\leq 0$ of
$k$})
%\left( \frac{\pa
%(\mL^n)_{\leq 0}}{\pa t_n} + \frac{\pa}{\pa t_n} \left(\mL
%-\sum_{m=1}^{\infty} \frac{1}{m}\frac{\pa v_m}{\pa x}
%\mL^{-m}\right) \right)
\end{align*}
and
\begin{align}\label{identity3}
&-\frac{\pa \M}{\pa \mL}\frac{\pa \mL}{\pa k} \frac{\pa \mL}{\pa
t_n} +\frac{\pa\mL}{\pa k}\left(\frac{\pa \M}{\pa \mL}\frac{\pa
\mL}{\pa t_n} + \frac{\pa \M}{\pa t_n}\Bigr\vert_{\mL
\fixed}\right)\\
=&\frac{\pa \mL}{\pa k} \left( n\mL^{n-1} + \sum_{m=1}^{\infty}
\frac{\pa v_m}{\pa t_n} \mL^{-m-1}\right)\nonumber\\
=&\frac{\pa (\mL^n)_{>0}}{\pa k} +(\text{powers $< 0$ of
$k$}).\nonumber
\end{align}
Comparing powers with the right hand side, we have
\begin{align*}
\ma{\frac{\pa \M}{\pa x}}{-\frac{\pa \mL}{\pa x}}{-\frac{\pa
\M}{\pa k}}{\frac{\pa \mL}{\pa k}}\ve{\frac{\pa \mL}{\pa
t_n}}{\frac{\pa \M}{\pa t_n}}= \ve{-\frac{\pa (\mL^n)_{>0}}{\pa
x}}{\frac{\pa (\mL^n)_{>0}}{\pa k}},
\end{align*}
or
\begin{align*}
\ve{\frac{\pa \mL}{\pa t_n}}{\frac{\pa \M}{\pa t_n}}=
\ma{\frac{\pa \mL}{\pa k}}{\frac{\pa \mL}{\pa x}}{\frac{\pa
\M}{\pa k}}{\frac{\pa \M}{\pa x}}\ve{-\frac{\pa (\mL^n)_{>0}}{\pa
x}}{\frac{\pa (\mL^n)_{>0}}{\pa k}},
\end{align*}
i.e.
\begin{align}\label{equ1}
\frac{\pa \mL}{\pa t_n} =\{ (\mL^n)_{>0}, \mL\} \hspace{1cm}
\text{and} \hspace{1cm}\frac{\pa \M}{\pa t_n} =\{ (\mL^n)_{>0},
\M\}.
\end{align}
Since the negative powers part of \eqref{identity3} vanishes,
rewriting the last equality in \eqref{identity3}, we have
\begin{align*}
\frac{\pa (\mL^n)_{>0}}{\pa k} = \frac{\pa}{\pa k}\left( \mL^n
-\sum_{m=1}^{\infty} \frac{1}{m} \frac{\pa v_m}{\pa t_n}
\mL^{-m}\right).
\end{align*}
Integrating with respect to $k$, we get
\begin{align}\label{identity4}
(\mL^n)_{>0}=\mL^n -\sum_{m=1}^{\infty} \frac{1}{m} \frac{\pa
v_m}{\pa t_n} \mL^{-m}-(\mL^n)_0.
\end{align}

Taking derivative of \eqref{def} with respect to $s$, as the case
of $t_n$, we have
\begin{align*}
\ma{\frac{\pa \M}{\pa x}}{-\frac{\pa \mL}{\pa x}}{-\frac{\pa
\M}{\pa k}}{\frac{\pa \mL}{\pa k}}\ve{\frac{\pa \mL}{\pa
s}}{\frac{\pa \M}{\pa s}} =\ma{\frac{\pa \wh{\M}}{\pa
x}}{-\frac{\pa \wh{\mL}}{\pa x}}{-\frac{\pa \wh{\M}}{\pa
k}}{\frac{\pa \wh{\mL}}{\pa k}}\ve{\frac{\pa \wh{\mL}}{\pa
s}}{\frac{\pa \wh{\M}}{\pa s}}.
\end{align*}
Now we have
\begin{align*}
\frac{\pa \M}{\pa x}\frac{\pa \mL}{\pa s}-\frac{\pa \mL}{\pa
x}\frac{\pa \M}{\pa s}=&\left(\frac{\pa \M}{\pa \mL}\frac{\pa
\mL}{\pa x}+ \frac{\pa \M}{\pa x}\Bigr\vert_{\mL
\fixed}\right)\frac{\pa \mL}{\pa s}-\frac{\pa \mL}{\pa
x}\left(\frac{\pa \M}{\pa \mL}\frac{\pa \mL}{\pa s} + \frac{\pa
\M}{\pa s} \Bigr\vert_{\mL \fixed} \right)\\
=&\left(1+ \sum_{m=1}^{\infty} \frac{\pa v_m}{\pa x}
\mL^{-m-1}\right) \frac{\pa \mL}{\pa s} -\frac{\pa \mL}{\pa x}
\left(\frac{N}{\mL} + \sum_{m=1}^{\infty} \frac{\pa v_m}{\pa s}
\mL^{-m-1} \right)\\
=&\frac{\pa}{\pa s}\left( \mL -\sum_{m=1}^{\infty} \frac{1}{m}
\frac{\pa v_m}{\pa x} \mL^{-m} \right)-\frac{\pa }{\pa x}
\left(\log \mL^N -\sum_{m=1}^{\infty} \frac{1}{m}\frac{\pa
v_m}{\pa s} \mL^{-m}\right).
\end{align*}
From \eqref{identity4}, the term in the first bracket is
\[
(\mL)_{\geq 0} = k+ u_1.
\]
We denote the term in the second bracket as \[ \log \mathcal{Q}
=\log \mL^N -\sum_{m=1}^{\infty} \frac{1}{m}\frac{\pa v_m}{\pa s}
\mL^{-m}.
\]
 Hence
\begin{align*}
\frac{\pa \M}{\pa x}\frac{\pa \mL}{\pa s}-\frac{\pa \mL}{\pa
x}\frac{\pa \M}{\pa s}=& \frac{\pa u_1}{\pa s} -\frac{\pa \log
\mathcal{Q}}{\pa x}.
\end{align*}
On the other hand,
\begin{align*}
-\frac{\pa \M}{\pa k} \frac{\pa \mL}{\pa s} + \frac{\pa \mL}{\pa
k}\frac{\pa \M}{\pa s} =& -\frac{\pa \M}{\pa \mL}\frac{\pa
\mL}{\pa k} \frac{\pa \mL}{\pa s} + \frac{\pa \mL}{\pa
k}\left(\frac{\pa \M}{\pa \mL}\frac{\pa \mL}{\pa s}+
\frac{N}{\mL}+\sum_{m=1}^{\infty}\frac{\pa v_m}{\pa
s} \mL^{-m-1}\right)\\
=&\frac{\pa \log \mathcal{Q}}{\pa k}.
\end{align*}
Hence we have
\begin{align}\label{identity6}
\ma{\frac{\pa \wh{\M}}{\pa x}}{-\frac{\pa \wh{\mL}}{\pa
x}}{-\frac{\pa \wh{\M}}{\pa k}}{\frac{\pa \wh{\mL}}{\pa
k}}\ve{\frac{\pa \wh{\mL}}{\pa s}}{\frac{\pa \wh{\M}}{\pa
s}}=\ma{\frac{\pa \M}{\pa x}}{-\frac{\pa \mL}{\pa x}}{-\frac{\pa
\M}{\pa k}}{\frac{\pa \mL}{\pa k}}\ve{\frac{\pa \mL}{\pa
s}}{\frac{\pa \M}{\pa s}} =\ve{\frac{\pa u_1}{\pa s}-\frac{\pa
\log \mathcal{Q}}{\pa x}}{\frac{\pa \log \mathcal{Q}}{\pa k}}.
\end{align}
On the other hand, since $\{\wh{\mL}, \wh{\M}\}=1$, we have the
identity
\begin{align*}
\ma{\frac{\pa \wh{\M}}{\pa x}}{-\frac{\pa \wh{\mL}}{\pa
x}}{-\frac{\pa \wh{\M}}{\pa k}}{\frac{\pa \wh{\mL}}{\pa k}}
\ve{\{\log \mP, \wh{\mL}\}}{\{\log\mP, \wh{\M}\}}=\ve{-\frac{\pa
\log \mathcal{P}}{\pa x}}{\frac{\pa \log \mathcal{P}}{\pa k}}.
\end{align*}
Hence
\begin{align*}
\ma{\frac{\pa \wh{\M}}{\pa x}}{-\frac{\pa \wh{\mL}}{\pa
x}}{-\frac{\pa \wh{\M}}{\pa k}}{\frac{\pa \wh{\mL}}{\pa k}}
\ve{\frac{\pa \wh{\mL}}{\pa s}-\{\log \mP, \wh{\mL}\}}{\frac{\pa
\wh{\M}}{\pa s}-\{\log\mP, \wh{\M}\}}=\ve{\frac{\pa u_1}{\pa
s}-\frac{\pa}{\pa x} \log \frac{\mathcal{Q}}{\mP}}{\frac{\pa}{\pa
k} \log \frac{\mathcal{Q}}{\mP}}.
\end{align*}
But by the conditions given on $\wh{\mL}$ and $\wh{\M}$, we have
\begin{align*}
\frac{\pa \wh{\M}}{\pa x} \left(\frac{\pa \wh{\mL}}{\pa s}-\{\log
\mP, \wh{\mL}\}\right) - \frac{\pa \wh{\mL}}{\pa x}\left(\frac{\pa
\wh{\M}}{\pa s}-\{\log\mP, \wh{\M}\}\right) = \left( \text{ powers
$\geq 1$ of $k$}\right),\\
-\frac{\pa \wh{\M}}{\pa k} \left(\frac{\pa \wh{\mL}}{\pa s}-\{\log
\mP, \wh{\mL}\}\right)+ \frac{\pa \wh{\mL}}{\pa k}\left(\frac{\pa
\wh{\M}}{\pa s}-\{\log\mP, \wh{\M}\}\right)= \left( \text{ powers
$\geq -1$ of $k$}\right).
\end{align*}
But by the normalization on $Q$ and $P$, we have
\begin{align*}
\frac{\pa u_1}{\pa s}-\frac{\pa}{\pa x} \log
\frac{\mathcal{Q}}{\mP} &=\left( \text{ powers $\leq 0$ of
$k$}\right),\\
\frac{\pa}{\pa k} \log \frac{\mathcal{Q}}{\mP}&=\left( \text{
powers $\leq -2$ of $k$}\right).
\end{align*}
Comparing powers, all the terms vanish, i.e.
\begin{align*}
\frac{\pa u_1}{\pa s}=0, \hspace{1cm} \frac{\pa}{\pa x} \log
\frac{\mathcal{Q}}{\mP}=0, \hspace{1cm} \frac{\pa}{\pa k} \log
\frac{\mathcal{Q}}{\mP}=0.
\end{align*}
Since $\log \frac{\mathcal{Q}}{\mP}$ contains powers $\leq -1$ of
$k$, it has to vanish identically. In other words, we have $\mP=
\mathcal{Q}$ and
\begin{align}\label{identity5}
 \log \mathcal{P}
=\log \mL^N -\sum_{m=1}^{\infty} \frac{1}{m}\frac{\pa v_m}{\pa s}
\mL^{-m}.
\end{align}
From \eqref{identity6}, we have
\begin{align*}
\ma{\frac{\pa \M}{\pa x}}{-\frac{\pa \mL}{\pa x}}{-\frac{\pa
\M}{\pa k}}{\frac{\pa \mL}{\pa k}}\ve{\frac{\pa \mL}{\pa
s}}{\frac{\pa \M}{\pa s}} =\ve{-\frac{\pa \log \mathcal{P}}{\pa
x}}{\frac{\pa \log \mathcal{P}}{\pa k}}.
\end{align*}
or
\begin{align}\label{identity7}
\frac{\pa \mL}{\pa s} =\{\log \mP, \mL\} , \hspace{1.5cm}\frac{\pa
\M}{\pa s} =\{\log \mP, \M\}.
\end{align}

 Finally, from \eqref{identity4} \eqref{identity5}, \eqref{equ1} and \eqref{identity7}, we get
 \begin{align*}
 &\frac{\pa \log \mP}{\pa t_n}-
 \frac{\pa (\mL^n)_{\geq 0} }{\pa s} + \{\log\mP, (\mL^n)_{>0}\}\\
= &\frac{\pa}{\pa t_n} \left(\log \mL^N -\sum_{m=1}^{\infty}
\frac{1}{m}\frac{\pa v_m}{\pa s} \mL^{-m}\right)-\frac{\pa }{\pa
s}\left(\mL^n -\sum_{m=1}^{\infty} \frac{1}{m}
\frac{\pa v_m}{\pa t_n} \mL^{-m}\right)\\
&+\frac{\pa \log \mP}{\pa k}
 \frac{\pa (\mL^n)_{>0}}{\pa x} - \frac{\pa \log \mP}{\pa x}
 \frac{\pa (\mL^n)_{>0}}{\pa k} \\
=&\frac{\pa \log \mP}{\pa \mL}\frac{\pa \mL}{\pa t_n}-\frac{\pa
(\mL^n)_{>0}}{\pa \mL}\frac{\pa \mL}{\pa s}+\frac{\pa \log
\mP}{\pa \mL}\frac{\pa \mL}{\pa k} \frac{\pa (\mL^n)_{>0}}{\pa x}
- \frac{\pa \log \mP}{\pa x}\frac{\pa (\mL^n)_{>0}}{\pa
\mL}\frac{\pa \mL}{\pa k}\\
=&\frac{\pa \log \mP}{\pa \mL}\left(\frac{\pa \mL}{\pa t_n} +
\frac{\pa (\mL^n)_{>0}}{\pa x}\frac{\pa \mL}{\pa
k}\right)-\frac{\pa (\mL^n)_{>0}}{\pa \mL} \left( \frac{\pa
\mL}{\pa s} +
\frac{\pa \log \mP}{\pa x}\frac{\pa \mL}{\pa k}\right)  \\
=&\frac{\pa \log \mP}{\pa \mL}\frac{\pa (\mL^n)_{>0}}{\pa
k}\frac{\pa \mL}{\pa x}-\frac{\pa (\mL^n)_{>0}}{\pa \mL}\frac{\pa
\log \mP}{\pa k}\frac{\pa \mL}{\pa x}=0.
\end{align*}
The fact that $\M$ is the corresponding Orlov function follows
from the characterization of Orlov functions in Proposition
\ref{orlov}.

\end{proof}
Conversely, we have
\begin{proposition}
If $(\mL,\mP, \M)$ is a solution of dcmKP hierarchy, then there
exists a pair of functions $(f,g)$ such that $\{ f,g\}=1$ and
satisfies \eqref{identity8} in Proposition \ref{twistor}
\end{proposition}
\begin{proof}
We let
\[
f(k,x) = e^{-\ad \varphi (s=0, t_n=0)} k, \hspace{1cm} g(k,x) =
e^{-\ad \varphi (s=0, t_n=0)}x.
\]
Then obviously $\{f,g\}=1$. The proof that $f, g$ satisfies the
first two conditions in \eqref{identity8} is standard (see
Proposition 1.5.2 in \cite{TT1}). Since $\mL, \M$ satisfis
\eqref{identity7}, the other two conditions follows from the
identities
\begin{align*}
\frac{\pa f(\mL, \M)}{\pa s} = \{ \log \mP, f(\mL, \M)\},
\hspace{1cm}\frac{\pa g(\mL, \M)}{\pa s} = \{ \log \mP, g(\mL,
\M)\}
\end{align*}
and the first two conditions.
\end{proof}
\section{$w_{1+\infty}$ symmetry}
We consider the $w_{1+\infty}$ action on the space of solutions of
the dcmKP hierarchy. Explicitly speaking, we define an
infinitesimal deformation of $(f,g)$ by a Hamiltonian vector
field,
\[
(f,g) \longrightarrow (f,g) \circ \exp(-\vep \ad F),
\]
and the associated deformation
\begin{align*}
(\mL,\mP, \M) \longrightarrow (\mL(\vep),\mP(\vep), \M(\vep))
\end{align*}
of the solution of dcmKP hierarchy. Here $\ad F$ is regarded as a
Hamiltonian vector field
\[
\ad F =\frac{\pa F}{\pa k}\frac{\pa}{\pa x}-\frac{\pa F}{\pa
x}\frac{\pa}{\pa k},
\]
and $\vep$ is an infinitesimal parameter. The infinitesimal
symmetry is the first order coefficient $\delta_F(\cdot)$ in the
$\vep$-expansion:
\[
\mL(\vep) = \mL + \vep\delta_F\mL + O(\vep^2),
\hspace{1cm}\M(\vep) = \M + \vep\delta_F\M + O(\vep^2).
\]
By definition, if $G$ is a function of $\mL$ and $\M$, then
\[
\delta_F G(\mL, \M)= \frac{\pa G}{\pa \mL} \delta_F \mL +
\frac{\pa G}{\pa \M} \delta_F \M,
\]
while the independent variables are invariant : $\delta_F  t=0$.

The infinitesimal symmetries of the dcmKP hierarchy is given by
the following propositions.
\begin{proposition}\label{symmetry1}
The infinitesimal symmetry of $\mL$, $\M$ and $\mP$ are given by
\begin{align*}
\delta_F \mL =& \{ F(\mL, \M)_{\leq 0}, \mL\}, \hspace{1cm}
\delta_F \M = \{F(\mL, \M)_{\leq 0}, \M\},\\
\delta_F \log \mP = &\frac{\pa F(\mL, \M)_{<0}}{\pa s}+\{ F(\mL,
\M)_{\leq 0},\log \mP\}.
\end{align*}
\end{proposition}
\begin{proof}
By definition, the twistor data $(f,g)$ is deformed as
\begin{align*}
&\left(f_{\vep}(k,x) , g_{\vep}(k,x)\right) = \left( e^{-\vep \ad
F} f(k,x), e^{-\vep\ad F} g(k,x)\right) \\
=& \left(f + \vep\left(\frac{\pa f}{\pa k}\frac{\pa F}{\pa
x}-\frac{\pa f}{\pa x}\frac{\pa F}{\pa k}\right), g +
\vep\left(\frac{\pa g}{\pa k}\frac{\pa F}{\pa x}-\frac{\pa g}{\pa
x}\frac{\pa F}{\pa k}\right)\right)+O(\vep^2).
\end{align*}
Hence, from
\begin{align*}
\wh{\mL}(\vep) =
f_{\vep}\left(\mL(\vep),\M(\vep)\right),\hspace{1cm} \wh{\M}(\vep)
= g_{\vep}\left(\mL(\vep),\M(\vep)\right),
\end{align*}
we read off the coefficients of $\vep$:
\begin{align*}
\ve{\delta_F \wh{\mL}}{\delta_F \wh{\M}} = \ma{\frac{\pa f}{\pa
\mL}}{\frac{\pa f}{\pa \M}}{\frac{\pa g}{\pa \mL}}{\frac{\pa
g}{\pa \M}} \ve{ \delta_F\mL + \frac{\pa F}{\pa \M}}{\delta_F \M -
\frac{\pa F}{\pa \mL}}.
\end{align*}
Now as in the proof of Proposition \ref{twistor}, we have
\begin{align*}
\ma{\frac{\pa \wh{\M}}{\pa x}}{-\frac{\pa \wh{\mL}}{\pa
x}}{-\frac{\pa \wh{\M}}{\pa k}}{\frac{\pa \wh{\mL}}{\pa k}}
\ve{\delta_F \wh{\mL}}{\delta_F \wh{\M}} = \ma{\frac{\pa \M}{\pa
x}}{-\frac{\pa \mL}{\pa x}}{-\frac{\pa \M}{\pa k}}{\frac{\pa
\mL}{\pa k}}\ve{ \delta_F\mL + \frac{\pa F}{\pa \M}}{\delta_F \M -
\frac{\pa F}{\pa \mL}}.
\end{align*}
Comparing powers of $k$, we have
\begin{align*}
\frac{\pa \M}{\pa x}\left(\delta_F\mL + \frac{\pa F}{\pa
\M}\right) -\frac{\pa \mL}{\pa x}\left(\delta_F \M - \frac{\pa
F}{\pa
\mL}\right) = \left( \text{powers $>0$ of $k$}\right),\\
-\frac{\pa \M}{\pa k}\left(\delta_F\mL + \frac{\pa F}{\pa
\M}\right) +\frac{\pa \mL}{\pa k}\left(\delta_F \M - \frac{\pa
F}{\pa \mL}\right) = \left( \text{powers $\geq 0$ of $k$}\right).
\end{align*}
As in the proof of Proposition \ref{twistor}, these give
\begin{align*}
\frac{\pa \M}{\pa x}\delta_F\mL-\frac{\pa \mL}{\pa x}\delta_F \M
&= -\frac{\pa }{\pa x} F(\mL, \M)_{\leq 0},
\\
-\frac{\pa \M}{\pa k}\delta_F\mL+\frac{\pa \mL}{\pa k}\delta_F \M
&= \frac{\pa }{\pa k} F(\mL, \M)_{< 0}= \frac{\pa }{\pa k} F(\mL,
\M)_{\leq 0}.
\end{align*}
Hence
\begin{align*}
\delta_F \mL =& \{ F(\mL, \M)_{\leq 0}, \mL\}, \hspace{1cm}
\delta_F \M = \{F(\mL, \M)_{\leq 0}, \M\}.
\end{align*}

Now from
\begin{align*}
\frac{\pa \mL(\vep)}{\pa s} =\{ \log \mP(\vep), \mL(\vep)\} ,
\end{align*}
we have
\begin{align*}
\frac{\pa}{\pa s} \delta_F \mL =\{ \delta_F \log \mP, \mL\}+\{\log
\mP, \delta_F \mL\}.
\end{align*}
Using the results above, we have
\begin{align*}
&\{ \delta_F \log \mP, \mL\} =\{ \frac{\pa}{\pa s} F(\mL,
\M)_{\leq 0}, \mL\} + \{F(\mL, \M)_{\leq 0}, \frac{\pa \mL}{\pa
s}\} -\{ \log \mP, \{F(\mL, \M)_{\leq 0}, \mL\}\}\\
=&\{ \frac{\pa}{\pa s} F(\mL, \M)_{\leq 0}, \mL\} + \{F(\mL,
\M)_{\leq 0}, \{ \log \mP, \mL\} \} +\{ \log \mP, \{ \mL, F(\mL,
\M)_{\leq 0}\}\}
\\=&\{ \frac{\pa}{\pa s} F(\mL,
\M)_{\leq 0}, \mL\}+\{\{F(\mL, \M)_{\leq 0}, \log \mP\} ,\mL\}.
\end{align*}
In other words,
\begin{align*}
\{\delta_F \log \mP -\frac{\pa}{\pa s} F(\mL, \M)_{\leq
0}-\{F(\mL, \M)_{\leq 0}, \log \mP\},\mL\}=0.
\end{align*}
Similarly, if we use
\[
\frac{\pa \M(\vep)}{\pa s} =\{ \log \mP(\vep), \M(\vep)\},
\]
we have
\begin{align*}
\{\delta_F \log \mP -\frac{\pa}{\pa s} F(\mL, \M)_{\leq
0}-\{F(\mL, \M)_{\leq 0}, \log \mP\},\M\}=0.
\end{align*}
The next lemma implies that
\[
\delta_F \log \mP -\frac{\pa}{\pa s} F(\mL, \M)_{\leq 0}-\{F(\mL,
\M)_{\leq 0}, \log \mP\}
\]
is independent of $k$ and $x$. Comparing powers of $k$, we get
\[
\delta_F \log \mP =\frac{\pa}{\pa s} F(\mL, \M)_{< 0}+\{F(\mL,
\M)_{\leq 0}, \log \mP\}.
\]
\end{proof}

\begin{lemma}\label{lemma1}
If $\mathcal{A}$ is such that $\{\mathcal{A}, \mL\}=0$ and
$\{\mathcal{A}, \M\}=0$ , then  $\mathcal{A}$ is independent of
$k$ and $x$.
\end{lemma}
\begin{proof}
$\{\mathcal{A}, \mL\}=0$ and $\{\mathcal{A}, \M\}=0$ are
equivalent to
\begin{align*}
\ma{\frac{\pa \mL}{\pa x}}{-\frac{\pa \mL}{\pa k}}{\frac{\pa
\M}{\pa x}}{-\frac{\pa \M}{\pa k}}\ve{\frac{\pa \mathcal{A}}{\pa
k} }{\frac{\pa \mathcal{A}}{\pa x}}=\ve{0}{0}.
\end{align*}
Since $\{\mL, \M\}=1$, this implies that
\[
\frac{\pa \mathcal{A}}{\pa k} =0 , \hspace{1.5cm}\frac{\pa
\mathcal{A}}{\pa x} =0.
\]
In other words, $\mathcal{A}$ is independent of $k$ and $x$.
\end{proof}

\begin{proposition}\label{symmetry2}
The infinitesimal symmetries of the $v_n$'s are given by
\[
\delta_F v_n = -\Res F(\mL, \M) d_k\B_n.
\]
\end{proposition}
\begin{proof}
The proof follows the same line as Proposition 14 in \cite{TT1},
see the proof of Proposition \ref{duality}. We have
\begin{align*}
\delta_F \M\Bigr\vert_{\mL \fixed} = \sum_{n=1}^{\infty} \delta_F
v_n \mL^{-n-1}.
\end{align*}
This gives
\begin{align*}
\delta_F v_n =& \Res \mL^n \left(\delta_F \M-\frac{\pa \M}{\pa
\mL}\delta_F \mL\right)d_k\mL = \Res \mL^n d_kF(\mL,\M)_{\leq 0} \\
=& \Res (\mL^n)_{>0} d_k F(\mL, \M) = -\Res F(\mL, \M) d_k B_n.
\end{align*}
The second equality follows the same as the proof in Proposition
\ref{duality}.
\end{proof}

\begin{proposition}\label{symmetry3}

\noindent (1) The infinitesimal symmetry of the dressing function
$\varphi$ is determined (up to a function of $s$) by the relation
\begin{align*}
\nabla_{\delta_F, \varphi} \varphi= F(\mL, \M)_{\leq 0},
\end{align*}
or equivalently, by
\begin{align*}
\delta_F \varphi = \frac{\ad \varphi}{e^{\ad \varphi} -1}F(\mL,
\M)_{\leq 0}=\frac{T}{e^{\ad T}-1}\Biggr\vert_{T= \ad \varphi}
F(\mL, \M)_{\leq 0},
\end{align*}
where $\frac{T}{e^{\ad T}-1}$ is understood as a power series of
$T$.

\noindent (2)
 The infinitesimal symmetry of the function $\phi$ defined in
 Proposition \ref{phi} is given (up to a function of $s$) by
\begin{align*}
\delta_F \phi =- F(\mL, \M)_0
\end{align*}
\end{proposition}
\begin{proof}
First we proof (2). Let $\phi(\vep)$ be the function given by
Proposition \ref{phi} corresponding to $\mL(\vep)$. Since the
function $\phi(\vep)$ is defined up to functions of $s$, it is
sufficient to show that
\begin{align*}
\frac{\pa \delta_F \phi}{\pa t_n} = -\frac{\pa}{\pa t_n}F(\mL,
\M)_0= -\left(\frac{\pa}{\pa t_n} F(\mL, \M)\right)_0.
\end{align*}
Now
\begin{align*}
\frac{\pa \delta_F \phi }{\pa t_n}=&\delta_F\frac{\pa \phi}{\pa
t_n} =\delta_F (\mL^n)_0 = \Bigl(\{F(\mL, \M)_{\leq 0},
\mL^n\}\Bigr)_0
\\=&\Bigl(\{F(\mL, \M)_{\leq 0}, (\mL^n)_{>0}\}\Bigr)_0
=\Bigl(\{F(\mL, \M), (\mL^n)_{>0}\}\Bigr)_0\\
=&-\Bigl(\frac{\pa}{\pa t_n} F(\mL, \M)\Bigr)_0.
\end{align*}
This prove (2).

Now let $\varphi(\vep)$ be the dressing function of $\mL(\vep)$.
Comparing the coefficient of $\vep$ in
\begin{align*}
\mL(\vep) =e^{\ad \varphi(\vep)} k,\hspace{1cm} \M(\vep)=e^{\ad
\varphi(\vep)}\left(\sum_{n=1}^{\infty} nt_nk^{n-1} +x
+\frac{Ns}{k}\right),
\end{align*}
we have
\begin{align*}
\delta_F \mL = \{ \nabla_{\delta_F, \varphi} \varphi, \mL\},
\hspace{1cm} \delta_F \M = \{ \nabla_{\delta_F, \varphi} \varphi,
\M\}.
\end{align*}
Compare with Proposition \ref{symmetry1}, we have
\begin{align*}
\{ \nabla_{\delta_F, \varphi} \varphi-F(\mL, \M)_{\leq 0},
\mL\}=0, \hspace{1cm}\{ \nabla_{\delta_F, \varphi} \varphi-F(\mL,
\M)_{\leq 0}, \M\}=0.
\end{align*}
By Lemma \ref{lemma1}, this implies that
\begin{align}\label{ide1}
\nabla_{\delta_F, \varphi} \varphi-F(\mL, \M)_{\leq 0}
\end{align}
 is a constant independent of $x$. To determine this
constant, we have to find the coefficient of $k^0$ in
$\nabla_{\delta_F, \varphi} \varphi$. Writing
$\varphi=\sum_{n=0}^{\infty} \varphi_n k^{-n}$. Comparing the
$k^0$ term in the identity
\begin{align*}
\nabla_{t_n, \varphi} \varphi = -(\mL^n)_{\leq 0},
\end{align*}
we have $\frac{\pa \varphi_0}{\pa t_n} = -(\mL^n)_0$. From
Proposition \ref{phi}, this implies that up to a function of $s$,
$\varphi_0 = -\phi$. Hence the $k^0$ term in \eqref{ide1} is
\begin{align*}
\delta_F \varphi_0 - (F(\mL, \M))_0 = -\delta_F \phi -(F(\mL,
\M))_0.
\end{align*}
By the second part of the proposition we prove above, this
vanishes (up to a function of $s$). Hence we have the first part
of our proposition.
\end{proof}

\begin{proposition}\label{symmetry4}
The infinitesimal symmetry of the function $\Phi$ defined by
\eqref{con2} is given by
\begin{align*}
\delta_F \Phi = -\Res F(\mL, \M) d\log \mP.
\end{align*}
\end{proposition}
\begin{proof}
Again, since the function $\Phi(t)$ is defined up to a function of
$s$, it is sufficient to show that
\begin{align*}
\frac{\pa  \delta_F \Phi }{\pa t_n}= -\frac{\pa}{\pa t_n}
\Bigl(\Res F(\mL, \M) d\log \mP\Bigr).
\end{align*}
From \eqref{con2} and Proposition \ref{symmetry2},
\begin{align*}
&\frac{\pa  \delta_F \Phi }{\pa t_n}=\delta_F \frac{\pa \Phi}{\pa
t_n}= \delta_F \frac{\pa v_n}{\pa s}=\frac{\pa \delta_F v_n}{\pa
s}=-\frac{\pa}{\pa s} \Res F(\mL, \M) d_k(\mL^n)_{\geq 0}\\
=& -\Res \{\log \mP, F(\mL, \M)\} d_k\B_n + \Res \frac{\pa
(\mL^n)_{\geq 0}}{\pa s} d_kF(\mL, \M)\\
=&-\Res  \left(\frac{\pa \log \mP}{\pa k} \frac{\pa F(\mL,
\M)}{\pa x} - \frac{\pa \log \mP}{\pa x} \frac{\pa F(\mL, \M)}{\pa
k}\right) \frac{\pa \B_n}{\pa k}dk \\
&+ \Res \frac{\pa \log \mP}{\pa t_n} d_kF(\mL, \M) + \Res
\left(\frac{\pa \log \mP}{\pa k}\frac{\pa \B_n}{\pa x}-\frac{\pa
\log \mP}{\pa x} \frac{\pa \B_n}{\pa k}\right)\frac{\pa F(\mL,
\M)}{\pa k}dk\\
=&-\Res  \{\B_n, F(\mL, \M)\} d_k\log\mP +\Res \frac{\pa
\log\mP}{\pa t_n} d_kF(\mL, \M) \\
=& -\frac{\pa}{\pa t_n} \Res F(\mL, \M) d_k\log\mP.
\end{align*}
\end{proof}

\begin{remark}
Observe that in the special case $\mP =k$, $\phi = \Phi$. In fact
in this case
\begin{align*}
-\Res F(\mL, \M) d_k\log \mP = -\Res F(\mL, \M) d\log k=-F(\mL,
\M)_0.
\end{align*}
\end{remark}

\begin{remark}
Compare the infinitesimal symmetries of $\mL, \mP, v_n, \Phi,
\varphi$ given in the Propositions above with the $t_n$ flows of
this quantities, it suggests that the Hamiltonian vector field
generate by the function $F(k,x) = -k^n$ is equivalent to
$\frac{\pa}{\pa t_n}$. The discrepancy between the $t_n$ flow of
$\M$ with the infinitesimal symmetry of $\M$ when $F(k,x) =-k^n$
is because $\M$ depends explicitly on $t_n$, but we enforce
$\delta_F t_n =0$.
\end{remark}

It is worth notice that if $\mL$ is a solution to our dcmKP
hierarchy with $\mP=k$ and $(f,g)$ an associated twistor data,
then for any function $F(k,x)$, since
\begin{align*}
\frac{\pa F(\mL, \M)}{\pa s} = \{ \log \mP, F(\mL, \M)\} =
\frac{1}{k} \frac{\pa F(\mL, \M)}{\pa x},
\end{align*}
we have
\begin{align*}
\frac{\pa (F(\mL, \M))_{<0}}{\pa s} = \frac{1}{k} \frac{\pa
(F(\mL, \M))_{\leq 0}}{\pa x}.
\end{align*}
Hence from Proposition \ref{symmetry1}, we have
\begin{align*}
\delta_F \log\mP =0.
\end{align*}
In other words, the class of special solutions $\mP=k$ is stable
under the $w_{1+\infty}$ action.

\subsection{Symmetries extended to tau functions}

The above symmetries can be extended to tau functions as follows.
\begin{proposition}
The infinitesimal symmetries of the tau function is given (up to a
function of $s$) by
\begin{align*}
\delta_F \log \tau =-\Res F^x(\mL, \M) d_k\mL,
\end{align*}
where $F^x(k,x)$ is a primitive function of $F(k,x)$ normalized as
\[
F^x =\int_0^x F(k,y) dy.
\]
It is compatible with the flows:
\begin{align*}
\frac{\pa }{\pa t} \delta_F \log \tau = \delta_F \frac{\pa \log
\tau}{\pa t},
\end{align*}
where $t=s$ or $t_n$'s.
\end{proposition}
\begin{proof}
Let
\begin{align*}
F^x(\mL, \M) = \sum_{m \in \Z} F^m(t) \mL^{m},
\end{align*}
so that $\Res F^x(\mL, \M) d_k\mL=F_{-1}(t)$.  Then we have to
show that
\begin{align}\label{result}
-\frac{\pa F_{-1}(t)}{\pa t} = \delta_F  \frac{\pa \log \tau}{\pa
t},
\end{align}
for $t=s$ or $t_n$'s. Since the term $F_m(t)$'s come purely from
$\M$, we have
\begin{align*}
\sum_{m \in \Z} \frac{\pa F_m(t)}{\pa t}\mL^{m} = \frac{\pa
F^x}{\pa \M} \frac{\pa \M}{\pa t} \Bigr\vert_{\mL \fixed}= F(\mL,
\M)\frac{\pa \M}{\pa t} \Bigr\vert_{\mL \fixed}
\end{align*}
Hence as we have seen in the proof of Proposition \ref{duality},
\begin{align*}
\frac{\pa F_{-1}(t)}{\pa t}=\Res  F(\mL, \M)\frac{\pa \M}{\pa t}
\Bigr\vert_{\mL \fixed}d_k\mL=\Res F(\mL, \M) d_k\mathcal{A},
\end{align*}
where $\mathcal{A} =\B_n$ if $t =t_n$ and $\mathcal{A} = \log \mP$
if $t=s$. Since $\frac{\pa \log \tau}{\pa t}= v_n$ if $t=t_n$ and
$\frac{\pa \log \tau}{\pa t}= \Phi$ if $t=s$, it follows from
Propositions \ref{symmetry2} and \ref{symmetry4} that
\eqref{result} holds.
\end{proof}

The $w_{1+\infty}$ algebra structure is reflected as follows.
\begin{proposition}
For the functions $F_1(k,x)$ and $F_2(k,x)$, the infinitesimal
symmetries obey the commutation relations:
\begin{align*}
[\delta_{F_1}, \delta_{F_2}] \log \tau &=\delta_{\{F_1, F_2\}}\log
\tau +c(F_1, F_2),\\
[\delta_{F_1}, \delta_{F_2}]\mathcal{K} &= \delta_{\{F_1,
F_2\}}\mathcal{K}, \hspace{1cm} \mathcal{K} \in \{\mL, \M, \mP\},\\
[\delta_{F_1}, \delta_{F_2}]\varphi &=\delta_{\{F_1,
F_2\}}\varphi,\\
[\delta_{F_1}, \delta_{F_2}]\Phi &=\delta_{\{F_1,
F_2\}}\Phi,\\
[\delta_{F_1}, \delta_{F_2}]\phi &=\delta_{\{F_1,
F_2\}}\phi,\\
\end{align*}
where
\[
c(F_1, F_2) = \Res F_1(k,0)d_kF_2(k,0),
\]
a cocycle of the $w_{1+ \infty}$ algebra.
\end{proposition}
\begin{proof}
The proof of the first identity follows exactly the same as
Proposition 16 in \cite{TT1}. The second and fourth identities
follows from the first one and the consistency of the
infinitesimal symmetries of $\tau$ with time flows.

To prove the last identity, we have
\begin{align*}
&\delta_{F_1} \delta_{F_2} \phi = -\delta_{F_1} \Bigl( F_2(\mL,
\M)\Bigr)_0\\
&= -\Bigl(\{(F_1(\mL, \M))_{\leq 0}, F_2(\mL, \M)\}\Bigr)_0\\
&=-\left(\frac{\pa (F_1(\mL, \M))_{\leq 0}}{\pa k}\frac{\pa
F_2(\mL, \M)}{\pa x} -\frac{\pa (F_1(\mL, \M))_{\leq 0}}{\pa
x}\frac{\pa F_2(\mL, \M)}{\pa k}\right)_0\\
&=-\left(\frac{\pa (F_1(\mL, \M))_{\leq 0}}{\pa k}\frac{\pa
(F_2(\mL, \M))_{>0}}{\pa x} -\frac{\pa (F_1(\mL, \M))_{\leq
0}}{\pa x}\frac{\pa (F_2(\mL, \M))_{>0}}{\pa k}\right)_0
\end{align*}
Hence
\begin{align*}
&[ \delta_{F_1} , \delta_{F_2}] \Phi=\delta_{F_1} \delta_{F_2}
\Phi-\delta_{F_2} \delta_{F_1} \Phi\\
=&-\left(\frac{\pa (F_1(\mL, \M))_{\leq 0}}{\pa k}\frac{\pa
(F_2(\mL, \M))_{>0}}{\pa x} -\frac{\pa (F_1(\mL, \M))_{\leq
0}}{\pa x}\frac{\pa (F_2(\mL, \M))_{>0}}{\pa k}\right)_0 \\
&+\left(\frac{\pa (F_2(\mL, \M))_{\leq 0}}{\pa k}\frac{\pa
(F_1(\mL, \M))_{>0}}{\pa x} -\frac{\pa (F_2(\mL, \M))_{\leq
0}}{\pa x}\frac{\pa (F_1(\mL, \M))_{>0}}{\pa k}\right)_0\\
=&-\left(\frac{\pa F_1(\mL, \M)}{\pa k}\frac{\pa F_2(\mL, \M)}{\pa
x} -\frac{\pa F_1(\mL, \M)}{\pa x}\frac{\pa F_2(\mL, \M)}{\pa
k}\right)_0\\
 =&-\Bigl(
\{F_1, F_2\}(\mL, \M)\Bigr)_0= \delta_{\{F_1, F_2\}}\phi.
\end{align*}

For the third identity, observe that from the proof of Proposition
\ref{symmetry3}, we have $\varphi_0= -\phi$. From
\begin{align*}
e^{\ad \varphi} x &= x+ \sum_{n=1}^{\infty} v_n \mL^{-n-1}\\
&= x+ \sum_{n=1}^{\infty} (v_n + (\text{polynomials in $\{ u_1,
\ldots, u_{n-1}, v_1, \ldots, v_{n-1}\}$}))k^{-n-1},
\end{align*}
we can prove by induction that
\begin{align*}
v_n+ (\text{polynomials in $\{ u_1, \ldots, u_{n-1}, v_1, \ldots,
v_{n-1}\}$})\\ = -n \varphi_n + (\text{differential polynomials of
$\{\phi_0, \ldots, \phi_{n-1}\} $}),
\end{align*}
where the differential is taken with respect to $x$. Hence,
solving recursively, we have for $n \geq 1$,
\begin{align*}
\varphi_n = -\frac{v_n}{n} +(\text{differential polynomials of $\{
u_1, \ldots, u_{n-1}, v_1, \ldots, v_{n-1}$\}}).
\end{align*}
Hence the third identity follows from the second and fifth
identities and the consistency between the infinitesimal
symmetries and the $t$-flows.
\end{proof}
\section{Concluding remarks}
We have defined a dcmKP hierarchy which incorporate both the ones
defined by Kupershmidt, Chang and Tu \cite{Kuper, CT} and Takebe
\cite{T1}. Our motivation is to define a tau function for dmKP
hierarchy in Kupershmidt, Chang and Tu's version, so that it plays
the role of transition between the dToda hierarchy and dKP
hierarchy. From our point of view, a good tau function should
generate all the coefficients $u_n$'s in the formal power series
$\mL$. Hence, we find that it is necessary to introduce an extra
flow $s$. In our dcmKP hierarchy, the special case $\mP =k$ has a
tau function with the desired property, namely it generates the
coefficients $u_n$'s. For general $\mP$, it does not have this
property. However, our approach can be directly generalized to
several extra flows $s_1, s_2, \ldots, s_M$ with $M$ auxiliary
polynomials $\mP_1, \mP_2, \ldots, \mP_M$ to govern the flows. If
one of the $P_i$ is equal to $k$, then we will find a good tau
function. As a matter of fact, $\mP$ need not be a polynomial.
What we require is that the coefficient of the leading term is
one.

Our dcmKP hierarchy can also be considered as a quasiclassical
limit of a corresponding coupled modified KP (cmKP) hierarchy. It
will be interesting to establish the existence of a tau function
in the cmKP hierarchy, so that its quasiclassical limit is our
dispersionless tau function.

\vspace{0.3cm}

 \noindent \textbf{Acknowledgments.} We greatly
appreciate stimulating discussions with J.H. Chang, M.H. Tu and
J.C. Shaw. This work is partially supported by NSC grant NSC
91-2115-M-009-017.
\bibliographystyle{amsalpha}
\bibliography{mKP}

\end{document}